\documentclass[12pt]{article}
\pdfoutput=1
\usepackage{jheppub}
\usepackage[utf8]{inputenc}
\usepackage{amsmath}
\usepackage{amsfonts}
\usepackage{amssymb}
\usepackage{graphicx}
\usepackage[export]{adjustbox}
\newcommand{\eq}[1]{\begin{equation}#1\end{equation}}
\newcommand{\eqsplit}[1]{\begin{equation}\begin{split}#1\end{split}\end{equation}}
\newcommand{\ea}[1]{\begin{eqnarray}#1\end{eqnarray}}
\newcommand{\subeq}[1]{\begin{subequations}\begin{align}#1\end{align}\end{subequations}}

\newcommand{\bmat}{\left(\begin{array}}
\newcommand{\emat}{\end{array}\right)}

\def\yzero{\smash{\hbox{$y\kern-4pt\raise1pt\hbox{${}^\circ$}$}}}

\def\beq{\begin{equation}}
\def\eeq{\end{equation}}
\def\beqa{\begin{eqnarray}}
\def\eeqa{\end{eqnarray}}

\def\-{\hphantom{-}}
\def\ov{\overline}
\def\s2{\frac{1}{\sqrt2}}

\def\beq{\begin{equation}}
\def\eeq{\end{equation}}
\def\beqa{\begin{eqnarray}}
\def\eeqa{\end{eqnarray}}

\def\IF{\relax{\rm I\kern-.18em F}}
\def\II{\relax{\rm I\kern-.18em I}}

\def\Dsl{\,\raise.15ex\hbox{/}\mkern-13.5mu D} %this one can be subscripted
\def\IS{{\bf {S}}}

\def\IX{{\bf {X}}}

\def\NN{{\cal {N}}}

\def\lam{\lambda}

\catcode`\@=11   
\newdimen\@rotdimen
\newbox\@rotbox  

\def\@vspec#1{\special{ps:#1}}%  passes #1 verbatim to the output
\def\@rotstart#1{\@vspec{gsave currentpoint currentpoint translate
   #1 neg exch neg exch translate}}% #1 can be any origin-fixing transformation
\def\@rotfinish{\@vspec{currentpoint grestore moveto}}% gets back in synch 
\def\@rotr#1{\@rotdimen=\ht#1\advance\@rotdimen by\dp#1%
   \hbox to\@rotdimen{\hskip\ht#1\vbox to\wd#1{\@rotstart{90 rotate}%
   \box#1\vss}\hss}\@rotfinish}
\def\@rotl#1{\@rotdimen=\ht#1\advance\@rotdimen by\dp#1%
   \hbox to\@rotdimen{\vbox to\wd#1{\vskip\wd#1\@rotstart{270 rotate}%
   \box#1\vss}\hss}\@rotfinish}%
\def\@rotu#1{\@rotdimen=\ht#1\advance\@rotdimen by\dp#1%
   \hbox to\wd#1{\hskip\wd#1\vbox to\@rotdimen{\vskip\@rotdimen
   \@rotstart{-1 dup scale}\box#1\vss}\hss}\@rotfinish}%
\def\@rotf#1{\hbox to\wd#1{\hskip\wd#1\@rotstart{-1 1 scale}%
   \box#1\hss}\@rotfinish}%
\def\rotate{\@ifnextchar[{\@rotate}{\@rotate[l]}}
\def\@rotate[#1]#2{\setbox\@rotbox=\hbox{#2}\@nameuse{@rot#1}\@rotbox}

\catcode`\@=12
\begin{document}

%----------------------------------------------------------------------%
%  numbering equations with section number
%----------------------------------------------------------------------%
\makeatletter
\@addtoreset{equation}{section}
\makeatother
\renewcommand{\theequation}{\thesection.\arabic{equation}}
%----------------------------------------------------------------------%
%  title page
%----------------------------------------------------------------------%
\pagestyle{empty}
%\vspace*{1.0in}
\rightline{IFT-UAM/CSIC-21-86}
\vspace{1.2cm}
\begin{center}
\Large{\bf
Dynamical Cobordism\\
and Swampland Distance Conjectures}
\\[12mm] 
%}\\

\large{Ginevra Buratti, Jos\'e Calder\'on-Infante, \\Matilda Delgado,  Angel M. Uranga\\[4mm]}
\footnotesize{Instituto de F\'{\i}sica Te\'orica IFT-UAM/CSIC,\\[-0.3em] 
C/ Nicol\'as Cabrera 13-15, 
Campus de Cantoblanco, 28049 Madrid, Spain}\\ 
\footnotesize{\href{mailto:ginevra.buratti@uam.es}{ginevra.buratti@uam.es}, \href{mailto:j.calderon.infante@csic.es }{j.calderon.infante@csic.es}, \\ \href{mailto:matilda.delgado@uam.es}{matilda.delgado@uam.es},   \href{mailto:angel.uranga@csic.es}{angel.uranga@csic.es}}

\vspace*{5mm}

\small{\bf Abstract} \\%[5mm]
\end{center}
\begin{center}
\begin{minipage}[h]{\textwidth}
We consider spacetime-dependent solutions to string theory models with tadpoles for dynamical fields, arising from non-trivial scalar potentials. The solutions have necessarily finite extent in spacetime, and are capped off by boundaries at a finite distance, in a dynamical realization of the Cobordism Conjecture. We show that as the configuration approaches these cobordism walls of nothing, the scalar fields run off to infinite distance in moduli space, allowing to explore the implications of the Swampland Distance Conjecture. We uncover new interesting scaling relations linking the moduli space distance and the SDC tower scale to spacetime geometric quantities, such as the distance to the wall and the scalar curvature. We show that walls at which scalars remain at finite distance in moduli space correspond to domain walls separating different (but cobordant) theories/vacua; this still applies even if the scalars reach finite distance singularities in moduli space, such as conifold points.\\
We illustrate our ideas with explicit examples in massive IIA theory, M-theory on CY threefolds, and 10d non-supersymmetric strings. In 4d $\NN=1$ theories, our framework reproduces a recent proposal to explore the SDC using 4d string-like solutions.

\end{minipage}
\end{center}
\newpage
%----------------------------------------------------------------------%
%  Resetting of counters
%----------------------------------------------------------------------%
\setcounter{page}{1}
\pagestyle{plain}
\renewcommand{\thefootnote}{\arabic{footnote}}
\setcounter{footnote}{0}
%----------------------------------------------------------------------%
%  Paper begins
%----------------------------------------------------------------------%

\tableofcontents

\vspace*{1cm}

\newpage

\section{Introduction and Conclusions}

A remarkable proposal in the Swampland Program of quantum gravity constraints on effective field theories \cite{Vafa:2005ui} (see \cite{Brennan:2017rbf,Palti:2019pca,vanBeest:2021lhn} for reviews) is the Cobordism Conjecture \cite{McNamara:2019rup}, that is based on the expected absence of exact global symmetries in quantum gravity. In short, it states that any configuration in a consistent theory of quantum gravity should not carry any cobordism charge. In practice, it implies that any configuration in a consistent theory of quantum gravity should admit, at the topological level, the introduction of a boundary ending spacetime into nothing\footnote{This boundary may be dressed by additional defects, such as D-branes or O-planes in string setups, to absorb the relevant charges.}, in the sense of \cite{Witten:1981gj} (see \cite{Ooguri:2017njy,GarciaEtxebarria:2020xsr} for recent related discussions). Accordingly, we will refer to such boundaries as {\em walls of nothing}.
Equivalently, it implies that any two consistent theories of quantum gravity must admit, at the topological level, an interpolating configuration connecting them, as a generalized domain wall separating the two theories. We will refer to such configurations as {\em interpolating domain walls}.

The Cobordism Conjecture is topological in nature. However, it can lead to remarkable breakthroughs when supplemented by additional assumptions. For instance, the extra ingredient of supersymmetry of the theory (and possibly of its walls) has led to highly non-trivial constraints in lower dimensional theories, see e.g. \cite{Montero:2020icj,Tarazi:2021duw}. 

An important step forward in endowing cobordism walls with dynamics was taken in \cite{Buratti:2021yia}, in the study of theories with tadpoles for dynamical fields (dubbed {\em dynamical tadpoles}, as opposed to topological tadpoles, such as RR tadpoles, which lead to topological consistency conditions on the configuration\footnote{Note however that dynamical tadpoles were recently argued in \cite{Mininno:2020sdb} to relate to violation of swampland constraints of quantum gravity theories.}). These are ubiquitious in the presence of scalar potentials, and in particular in non-supersymmetric string models.
In theories with dynamical tadpoles the solutions to the equations of motion vary over the non-compact spacetime dimensions. Based on the behaviour of large classes of string models, it was proposed in \cite{Buratti:2021yia} that such spacetime-dependent running solutions must hit  cobordism walls of nothing at a finite distance $\Delta$ in spacetime\footnote{For related work on dynamical tadpoles in non-supersymmetric theories, see \cite{Dudas:2000ff,Blumenhagen:2000dc,Dudas:2002dg,Dudas:2004nd,Mourad:2016xbk,Basile:2018irz,Antonelli:2019nar,Basile:2020xwi}.} (as measured in the corresponding Einstein frame metric), scaling as $\Delta^{-n}\sim {\cal T}$ with the strength of the tadpole ${\cal T}$. These examples included holographic AdS$_5\times T^{1,1}$ compactifications with RR 3-form flux, type IIB 3-form flux compactifications, magnetized D-brane models, massive IIA theory, M-theory on K3 with $G_4$ flux, and the 10d non-supersymmetric $USp(32)$ string theory.
On the other hand, interpolating cobordism walls connecting different theories were not discussed. One of the motivations of this work is to fill this gap. 

We argue that, when a running solution in theories with dynamical tadpoles hits a wall, the behaviour of the configuration across the wall, and in particular the sharp distinction between interpolating domain walls and walls of nothing, is determined by the behaviour of scalar fields as one reaches the wall, via a remarkable correspondence: 

\smallskip

$\bullet$ When scalars remain at finite distance points in moduli space as one hits the wall, it corresponds to an interpolating domain wall, and the solution continues across it in spacetime (with jumps in quantities as determined by the wall properties);

$\bullet$ On the other hand, when the scalars run off to infinity in moduli space as one reaches the wall (recall, at a finite distance in spacetime), it corresponds to a wall of nothing, capping off spacetime beyond it. 

\smallskip

We also argue that scalars reaching singular points at finite distance in moduli space upon hitting the wall still define interpolating domain walls, rather than walls of nothing; hence, walls of nothing are not a consequence of general singularities in moduli space, but actually to those at infinity in moduli space. This suggests that, in the context of dynamical solutions\footnote{Note that, in setups with no dynamical tadpole, one can still have e.g. cobordism walls of nothing without scalars running off to infinity: for instance, 11d M-theory, which does not even have scalars,  admits walls of nothing defined by Horava-Witten boundaries; similar considerations may apply to potential theories with no moduli (or with all moduli stabilized at high enough scale).}, the walls of nothing of the Cobordism Conjecture are closely related to the Swampland Distance Conjecture\footnote{The status of the SDC in spacetime dependent running solutions was addressed in  \cite{Buratti:2018xjt}.}. We indeed find universal scaling relations between the (finite) distance to the wall in spacetime and the scale of the SDC tower \cite{Ooguri:2006in}. In addition, we uncover a universal scaling relation between the curvature scalar in running solutions and the SDC tower scale that is reminiscent of the Anti de Sitter Distance Conjecture (ADC) \cite{Lust:2019zwm}.

We illustrate these ideas in several large classes of string theory models, including massive IIA, and M-theory on CY threefolds.  Moreover, we also argue that our framework encompasses the recent discussion of EFT string solutions in 4d $\NN=1$ theories in \cite{Lanza:2021qsu} (see also \cite{Lanza:2020qmt}), where saxion moduli were shown to attain infinity in moduli space at the core of strings magnetically charged under the corresponding axion moduli. We show that EFT string solutions are the cobordism walls of nothing of $\IS^1$ compactifications of the  4d $\NN=1$ theory with certain axion fluxes on the $\IS^1$. Our scalings also relate to those between EFT string tensions and the SDC tower scale in \cite{Lanza:2021qsu}.

%$\bullet$ The discussion of the nonsusy $USp(32)$ theory and its running solution(s), should be included somewhere

The paper is organized as follows. In Section \ref{sec:iia} we present the main ideas in the explicit setup of running solutions in massive IIA theory, and their interplay with type I' solutions \cite{Polchinski:1995df}. In Section \ref{sec:mth} we carry out a similar discussion for M-theory on CY threefolds with $G_4$ flux (in Section \ref{sec:mth-general}) and their relation to strongly coupled heterotic strings \cite{Witten:1996mz}. In Section \ref{sec:mth-singular} we use it to discuss domain walls across singularities at finite distance in moduli space, following \cite{Greene:2000yb}. In Section \ref{sec:eft} we discuss the $\IS^1$ compactification of general 4d $\NN=1$ theories. In Section \ref{sec:circle} we introduce dynamical tadpoles from axion fluxes, whose running solutions hit walls of nothing at which saxions run off to infinity. In Section \ref{sec:eft-strings} we relate the discussion to the EFT strings of \cite{Lanza:2021qsu}. In Section \ref{sec:potentials} we discuss the moduli space distances in walls of nothing and interpolating walls in 4d $\NN=1$ theories with non-trivial superpotentials of the kind arising in flux compactifications. In Section \ref{sec:nonsusy} we discuss our proposal in non-supersymmetric string theories, in particular the 10d $USp(32)$ string. In Section \ref{sec:conclusions} we offer some final remarks and outlook. Appendix \ref{app:holographic} provides some observations on cobordism walls in holographic throats.

\section{Cobordism walls in massive IIA theory}
\label{sec:iia}

\subsection*{Walls of nothing and infinite moduli space distance}

In this section we consider different kinds of cobordism walls in massive IIA theory \cite{Romans:1985tz}, extending the analysis in \cite{Buratti:2021yia}. The Einstein frame 10d effective action for the relevant fields is
\beqa \label{eq:massive-typeIIA-action}
S_{10,E}\,=\,  \frac{1}{2\kappa^{\,2}}\int d^{10}x \,\sqrt{-G_E}\, \{\, [\, R-\frac 12(\partial \phi)^2\,]-\frac 12 e^{\frac 52\phi} F_0^{\,2}  -\frac 12 e^{\frac 12 \phi}(F_4)^2\,\}\, ,
\eeqa
where the Romans mass parameter is denoted by $F_0$ to suggest it is a 0-form field strength flux.
This theory is supersymmetric, but has a dilaton tadpole
\beqa
{\cal T}\,\sim \, e^{\frac 52\phi} F_0^2\ ,
\label{tadpole-massiveiia}
\eeqa
so the theory does not admit 10d maximally symmetric solutions. The solutions with maximal (super)symmetry are 1/2 BPS configurations with the dilaton depending on one coordinate $x^9$, closely related to that in \cite{Bergshoeff:1995vh}. In  conventions closer to \cite{Polchinski:1995df}, the Einstein frame metric and dilaton are
\beqa
(G_{E})_{MN}\, =\, Z(x^9)^{\frac 1{12}}\,\eta_{MN}\ ,\quad
e^{\phi}\,=\, Z(x^9)^{-\frac 56}\ , \quad{\rm with}\; \;Z(x^9)\sim -F_0\, x^9\ ,
\label{iia-bps-won}
\eeqa
where we have set some integration constant  to zero. The solution hits a singularity at $x^9=0$. The spacetime distance from a general position $x^9$ to the singularity is \cite{Buratti:2021yia}
\beqa
\Delta\,=\, \int_{x^9}^0 Z(x^9)^{\frac{1}{24} }\,dx^9\,\sim Z(x^9)^{\frac{25}{24}}\,F_0^{-1}\, \sim \,  F_0^{-1} e^{-\frac 54\phi}\, \sim\, {\cal T}^{-\frac 12}\, ,
\label{romans-spacetime-distance}
\eeqa
in agreement with the scaling relation $\Delta^{-2}\sim {\cal T}$, that was dubbed Finite Distance lesson in \cite{Buratti:2021yia}. Following the Dynamical Cobordism proposal therein, the singularity is resolved in string theory into a cobordism wall of nothing, defined by an O8-plane (possibly dressed with D8-branes to match the $F_0$ flux to be absorbed)\footnote{This imposes a swampland bound on the possible values of $F_0$ that are consistent in string theory.}, ending the direction $x^9$ as a boundary.

We now notice that, since $Z\to 0$ implies $\phi\to \infty$ as $x^9\to 0$, the dilaton runs off to infinity in moduli space as one hits the wall, as befits a wall of nothing from our discussion in the introduction. According to the SDC, there is an infinite tower of states becoming massless in this region, with a scale decaying exponentially with the moduli space distance $D$ as
\beqa
M_{\rm SDC}\sim e^{-\lambda D}\, ,
\eeqa
with some positive ${\cal{O}}(1)$ coefficient $\lambda$.

It is interesting to find a direct relation between these quantities and the spacetime distance to the wall. The distance in moduli space is given by $\phi=\sqrt{2}\,D$, as can be seen from the kinetic term for $\phi$ in \eqref{eq:massive-typeIIA-action}. From (\ref{romans-spacetime-distance}) we have
\beqa
%\Delta\sim F_0^{-1} e^{-\frac 54 d}\quad ,\quad M_{\rm SDC}\sim (F_0\Delta)^{\frac 45 \lambda}
\Delta\sim e^{-\frac{5}{2\sqrt{2}} D}\quad ,\quad M_{\rm SDC}\sim \Delta^{\frac{2\sqrt{2}}{5} \lambda}\, .
\eeqa
Hence the SDC tower scale goes to zero with the distance to the wall with a power-like scaling.

It is a natural question to ask if this tower of states becomes light in the actual dynamical configuration (rather than in the adiabatic framework of the standard formulation of the SDC). In this particular setup, the SDC tower corresponds to D0-branes which end up triggering the decompactification of the M-theory eleventh dimension. In the dynamical solution, there are a finite number of extra massless states, responsible for the enhancement of the perturbative open string gauge group to the exceptional symmetries which are known to arise from the heterotic dual theory \cite{Polchinski:1995df} (see also \cite{Seiberg:1996bd}). On the other hand, there is no signal of an infinite tower of states becoming massless simultaneously. The appearance of the SDC in the dynamical context has thus different implications as compared with the usual adiabatic formulation.

Let us now turn to another novel, and tantalizing, scaling. The scalar curvature for the running solution reads
\begin{equation}
    |R| \sim (-x^9)^{-\frac{25}{12}}  \sim e^{\frac{5}{\sqrt{2}}D} \, .
\end{equation}
Using this, we can write the SDC tower scale in terms of the scalar curvature as
\begin{equation}
    M_{\rm SDC} \sim e^{-\lambda D} \sim |R|^{-\frac{\sqrt{2}}{5}\lambda} \, .
\end{equation}
This scaling is highly reminiscent of the Anti de Sitter Distance Conjecture (ADC) of \cite{Lust:2019zwm}\footnote{It is possible that the result is ultimately linked to the generalized distance conjectures in \cite{Lust:2019zwm}; we leave this as an open question for future work.}, even though the setup under consideration is very different.\footnote{In contrast to the ADC, that considers the limit of vanishing curvature of a family of AdS vacua, in our setup the scalar curvature blows up as the singularity is approached. However, we do find a power-like scaling similar to the ADC one.} Note however that, as in the ADC, it signals a failure of the decoupling of scales, and hence a breakdown of the effective field theory near the wall of nothing. This fits nicely with our observation that the wall can only be microscopically defined in the UV complete theory, and works as a boundary condition defect at the level of the effective theory.

\subsection*{Interpolating domain walls}

There is a well known generalization of the above solutions, which involves the inclusion of D8-branes acting as interpolating domain walls across which $F_0$ jumps by one unit. The general solution of this kind is provided by (\ref{iia-bps-won}) with a piecewise constant $F_0$ and a piecewise continuous function $Z$ \cite{Polchinski:1995df}.

The D8-brane domain walls are thus (a very simple realization of) cobordism domain walls interpolating between different Romans IIA theories (differing just in their mass parameter). The point we would like to emphasize is that, since $Z$ remains finite across them, the dilaton remains at finite distance in moduli space, as befits interpolating domain walls from our discussion in the introduction.

\section{Cobordism walls in M-theory on CY3}
\label{sec:mth}

In this section we recall results from the literature on the strong coupling limit of the  heterotic string, also known as heterotic M-theory \cite{Witten:1996mz,Lukas:1998yy,Lukas:1998tt,Lukas:1998uy} (see \cite{Ovrut:1999xu,Ovrut:2002hi} for review and additional references). They provide straightforward realizations of the different kinds of cobordism walls in M-theory compactifications on CY threefolds. The discussion generalizes that in \cite{Buratti:2021yia}, and allows to study the behaviour at singular points at finite distance in moduli space, in particular flops at conifold points.

\subsection{M-theory on CY3 with $G_4$ flux}
\label{sec:mth-general}

We consider M-theory on a CY threefold $\IX$, with $G_4$ field strength fluxes on 4-cycles. For later convenience, we follow the presentation in \cite{Greene:2000yb}. We introduce dual basis of 2- and 4-cycles  $C^i\in H_2(\IX)$ and $D_i\in H_4(\IX)$, and define
\beqa
\int_{D_i} G_4\, =\, a_i \quad, \quad \int_{C^i} C_6={\tilde \lambda}^i ~.
\eeqa
We also denote by $b_i$ the 5d vector multiplet of real K\"ahler moduli, with the usual K\"ahler metric and the 5d $\NN=1$ prepotential
\begin{equation}
\label{eq:5}
G_{ij} =-\frac 12\frac{\partial^2}{\partial b_i \partial b_j} \ln {\cal{K}} ~,~~~~~
{\cal K} \equiv \frac 1{3!}d_{ijk} b^i b^j b^k ~,
\end{equation}
with $d_{ijk}$ being the triple intersection numbers of $\IX$. We have the familiar constraint ${\cal K}=1$ removing the overall modulus $V$, which lies in a hypermultiplet. 

The 5d effective action for these fields is 
\begin{eqnarray}
 S_5 &=& - \frac{M_{p,11}^{\,9}}{2} \,L^6\! \left[\int_{M_5}\sqrt{-g_5}\Big(R+G_{ij}(b)
         \partial_M b^i\partial^M b^j+ \frac{1}{2V^2}\partial_M V
         \partial^M V+\lambda ({\cal{K}} - 1)
         \Big)\right. \nonumber\\
      && \qquad\qquad
         \left. +\frac{1}{4V^2}  G^{ij}(b)a_i\wedge \star a_j
+d\tilde{\lam}^i
           \wedge a_i \right] -\sum_{n=0}^{N+1}
\alpha^{(n)}_i\int_{M_4^{(n)}}\big(\tilde{\lambda}^i
+\frac{b^i}{V}\sqrt{g_4}\big)~.  \quad\quad\quad
\label{S5_red}
\end{eqnarray}
Here ${\lambda}$ is a Lagrange multiplier, and $L$ the reference length scale of the Calabi-Yau. With hindsight, we include 4d localized terms which correspond to different walls in the theory, with induced 4d metric $g_4$. 

The $G_4$ fluxes $a_i$ induce dynamical tadpoles for the overall volume and the K\"ahler moduli $b_i$. There are 1/2 supersymmetric solutions running in one spacetime coordinate, denoted by $y$, with the structure
\begin{eqnarray}
\label{domainwallsoln}
ds_5^2&=&e^{2A}ds_4^2+e^{8A}dy^2 ~, \nonumber\\
V &=& e^{6A}\quad , \quad b^i = e^{-A} f^i ~, \nonumber\\
e^{3A}&=& \left( {1\over 3!} d_{ijk} f^i f^j f^k \right) ~, \nonumber\\
({d{\tilde \lambda}}^i)_{\mu\nu\rho\sigma} &=&\epsilon_{\mu\nu\rho\sigma} e^{-10A}
\left(-\partial_{11}b^i+2b^i\partial_{11}A\right)\, .
\end{eqnarray}
The whole solution is determined by a set of one-dimensional harmonic functions. They are given in terms of the local values of the $G_4$ fluxes, 
\begin{eqnarray}
\label{implicitsoln}
d_{ijk} f^j f^k = H_i ~, \quad \quad H_i &=& a_i y +c_i ~.
\end{eqnarray}
Here the $c_i$ are integration constants set to have continuity of the $H_i$, and hence of the $f_i$, across  the different interpolating domain walls in the system, which produce jumps as follows. Microscopically, the interpolating domain walls correspond to M5-branes wrapped on 2-cycles $[C]=\sum n_i C^i$, leading to jumps in the fluxes that in units of M5-brane charge are given by
\beqa
\Delta a_i = n_i ~.
\eeqa

Hence, interpolating domain walls maintain the theory at finite distance in moduli space. This is not the case for cobordism walls of nothing, which arise when $e^A\to 0$, and hence $V\to 0$, which sits at infinity in moduli space. This regime was already discussed (in the simpler setup of K3 compactifications) in \cite{Buratti:2021yia}, where the cobordism domain was argued to be given by a Horava-Witten boundary (dressed with suitable gauge bundle degrees of freedom, as required to absorb the local remaining $G_4$ flux), in agreement with the strong coupling singularity discussed in \cite{Witten:1996mz}. The wall appears at a finite spacetime distance $\Delta$ following the scaling $\Delta^{-2}\sim {\cal T}$ in \cite{Buratti:2021yia}. In what follows, we describe the scaling relations of the moduli space distance and the SDC tower at these walls of nothing. 

Since they are characterized by the vanishing of the overall volume of $\IX$, it is enough to follow the behaviour of $V$ and the discussion simplifies. Restriction to this sector amounts to setting all $f_i\equiv f$ in (\ref{domainwallsoln}), and all $H_i\equiv H$. Also, since the wall of nothing arises when $H\to 0$, we can take this location as $y=0$ and write
\ea{e^{2A}\sim H(y)\sim \alpha y\, .
}
Using the metric in (\ref{domainwallsoln}), the spacetime distance from a point $y>0$ is
\ea{\Delta\,=\int_0^y (\alpha y)^2 dy\, =\,\frac 13 \alpha^2 \,y^3\, .
\label{delta-cy3}
}
We are also interested in the traversed distance in moduli space $D$. Using the kinetic term in (\ref{S5_red}), the relevant integral to compute is
\ea{D\,=\,-\int \frac {1}{\sqrt{2}V}\frac{dV}{dy}dy \, .
}
Using $V\sim H^3$, we get as leading behavior near the singularity
\ea{D\simeq -\frac{3}{\sqrt{2}}\log y\,=\, -\frac 1{\sqrt{2}}\log \frac{3\Delta}{\alpha^2}\, ,}
where in the last equality we used (\ref{delta-cy3}).
This implies
\eq{\Delta\sim e^{-\sqrt{2}D}\, ,
}
and leads to a power-like scaling of the SDC tower mass 
\ea{M_{\rm SDC}\sim \Delta^{\frac{\lambda}{\sqrt{2}}}\, .
}
Computing the curvature scalar from (\ref{domainwallsoln}), we get
\eq{|R|\sim e^{2\sqrt{2}D}\, .
}
So the SDC tower scale can be expressed, in an ADC-like manner, as
\ea{M_{\rm SDC}\sim |R|^{-\frac{\lambda}{2\sqrt{2}}}\, .
}
We thus recover a similar behaviour to the examples in Section \ref{sec:iia}.

\subsection{Traveling across finite distance singularities in moduli space}
\label{sec:mth-singular}

The setup of M-theory on a CY3 $\IX$ allows to address the question of whether walls of nothing could arise at finite distance in moduli space, if the scalars hit a singular point in moduli space. This is actually not the case, as can be explicitly shown by following the analysis in \cite{Greene:2000yb} for flop transitions. 

Specifically, they considered the flop transition between two Calabi-Yau manifolds  with $(h_{1,1}, h_{2,1}) = (3, 243)$, in the setup of a CY3 compactification of the Horava-Witten theory, namely with two boundaries restricting the coordinate $y$ to an interval. In our more general setup, one may just focus on the dynamics in the bulk near the flop transition as one moves along $y$. Hence we are free to locate the flop transition point at $y=0$.

In terms of the K\"ahler moduli $t^i=V^{\frac 13} b_i$ of $\IX$, and changing to a more convenient basis
\begin{eqnarray}
  \label{eq:bb}
  t^1 = U\quad ,\quad 
  t^2 = T-\frac 12 U-W \quad ,\quad
  t^3 &=& W-U ~,
\end{eqnarray}
and similar (proper transforms under the flop) for ${\tilde \IX}$, the K\"{a}hler cones of $\IX$ and ${\tilde \IX}$ are defined by the regions
\begin{eqnarray}
  \label{eq:kc}
  \IX:&& W>U>0~,~T>\frac 12 U+W~, \\
{\tilde{\IX}}:&& U>W>0~,~T> \frac{3}{2} U\, .
\end{eqnarray}
This shows that the flop curve is $C_3$, and the area is $W-U$, changing sign across the flop.

Near the flop point $y=0$, the harmonic functions for the two CYs $\IX$ and ${\tilde \IX}$ have the form
\begin{eqnarray}
{\IX} \;{\rm at}\; y\leq 0 \quad\quad & & \quad\quad {\tilde \IX} \;{\rm at}\; y\geq 0\nonumber\\
  {H}_T = -18y+{k}_T \quad &, & \quad \tilde{H}_T = 18y+{k}_T ~,\nonumber\\
  {H}_U = -25y+{k}_0 \quad &, & \quad \tilde{H}_U =24y+ {k}_0 ~,\nonumber\\
  {H}_W = 6y+{k}_0 \quad &, & \quad \tilde{H}_W = -5y+{k}_0 ~.
\end{eqnarray}
Hence
\beqa
{\IX} \;{\rm at}\; y\leq 0 \quad\quad & & \quad\quad {\tilde \IX} \;{\rm at}\; y\geq 0\nonumber\\
H_{W-U}=31y  \quad &, & \quad {\tilde H}_{W-U}=-29y ~.
\eeqa
Even though the flop point is a singularity in moduli space, and despite the sign flip for $W-U$, the harmonic functions are continuous and the solution remains at finite distance in moduli space. This agrees with the picture that it corresponds to an interpolating domain wall. In fact, as discussed in \cite{Greene:2000yb}, the discontinuity in their slopes (and the related change in the $G_4$ fluxes) makes the flop point highly analogous to the above described interpolating domain walls associated to M5-branes.

The above example illustrates a further important aspect. It provides an explicit domain wall intepolating between two different (yet cobordant) topologies. It would be extremely interesting to extend this kind of analysis to other topology changing transitions, such as conifold transitions\footnote{For a proposal to realize conifold transitions dynamically in a time-dependent background, see \cite{Mohaupt:2004pr}.} \cite{Greene:1995hu}. This would allow for a further leap for the dynamical cobordism proposal, given that moduli spaces of all CY threefolds are expected to be connected by this kind of transitions \cite{Chiang:1995hi}.

We have thus established that physics at finite distance in moduli space gives rise to interpolating domain walls, rather than walls of nothing, even at singular points in moduli space. The implication is that the physics of walls of nothing is closely related to the behaviour near infinity in moduli space and hence to the SDC. In the following section we explore further instances of this correspondence in general 4d $\NN=1$ theories.

\section{$\IS^1$ compactification of 4d $\NN=1$ theories and EFT strings}
\label{sec:eft}

In this section we study a systematic way to explore infinity in moduli space in general 4d $\NN=1$ theories. This arises in a multitude of string theory constructions, ranging from heterotic CY compactifications to type II orientifolds on CY spaces \cite{Ibanez:2012zz}. Our key tool is an $\IS^1$ compactification to 3d with certain axion fluxes. We will show that the procedure secretly matches the construction of EFT strings in \cite{Lanza:2021qsu} (see also \cite{Lanza:2020qmt}). Actually, this correspondence was the original motivation for this paper.

\subsection{Cobordism walls in 4d $\NN=1$ theories on a circle}
\label{sec:circle}

We want to consider general 4d $\NN=1$ theories near infinity in moduli space. According to \cite{Grimm:2018ohb,Corvilain:2018lgw,Grimm:2019ixq}, the moduli space in this asymptotic regime is well approximated by a set of axion-saxion complex fields, with metric given by hyperbolic planes. We start discussing the single-field case, and sketch its multi-field generalization at the end of this section.

Consider a 4d $\NN=1$ theory with complex modulus $S=s+ia$, where $a$ is an axion of unit periodicity and $s$ its saxionic partner. We take a K\"ahler potential 
\beqa
K=-\frac 2{n^2} \log (S+{\bar S}) ~.
\eeqa
The 4d effective action is
\eqsplit{\label{4d-action}
S&=\frac{M^2_{P,4}}{2}\int d^4 x\sqrt{-g_4}\left\{ R_4-\frac{n^{-2}}{s^2}\left[\left(\partial s\right)^2+\left(\partial a\right)^2\right]\right\}\, ,\\
&=\frac{M^2_{P,4}}{2}\int d^4 x\sqrt{-g_4}\left\{ R_4-\left(\partial \phi\right)^2-e^{-2n\phi}\left(\partial a\right)^2\right\}\, ,
}
where in the last equation we have defined \(\phi= \frac{1}{n}\log{ns}\).

We now perform an $\IS^1$ compactification to 3d with the following ansatz for the metric\footnote{We omit the KK $U(1)$ because it will not be active in our discussion.} and the scalars
\beqa
ds_4^2&=&e^{-{\sqrt{2}\sigma}}ds_3^2+e^{\sqrt{2}\sigma} R_0^2  d\theta^2\, ,  \nonumber\\
\phi&=& \phi(x^\mu)\, , \qquad a=\frac{\theta}{2\pi}q+a(x^\mu)\, , \label{eq:compactification-ansatz}
\eeqa
where $x^\mu$ denote the 3d coordinates and $\theta\sim\theta+2\pi$ is a periodic coordinate. Regarding the axion as a 0-form gauge field, the ansatz for $a$ introduces $q$ units of its field strength flux (we dub it axion flux) on the $S^1$. We allowed for a general saxion profile to account for its backreaction, as we see next.

The dimensional reduction of the action \eqref{4d-action} gives (see e.g. \cite{PopeKK})
\eq{
S_3=\frac{M_{P,3}}{2}\int d^3 x\sqrt{-g_3} \left\{R_3-G_{ab}\partial_\mu\varphi^a\partial^\mu\varphi^b-V(\varphi)\right\}\, ,
}
where
\subeq{
G_{ab}\partial_\mu\varphi^a\partial^\mu\varphi^b&=\left(\partial \sigma\right)^2+\left(\partial \phi\right)^2+e^{-2n\phi}\left(\partial a\right)^2\, , \\
V(\varphi)&=e^{-2\sqrt{2}\sigma-2n\phi}\left(\frac{q}{2\pi R_0}\right)^2\, ,
}
and $M_{P,3}=2\pi R_0 M_{P,4} ^2$ is the 3d Planck mass. 

The last term in the 3d action corresponds to a dynamical tadpole for a linear combination of the saxion and the radion, induced by the axion flux. We thus look for running solutions of the 3d equations of motion. We focus on solutions with  constant axion in 3d $a(x^\mu)=0$, for which the equations of motion read
\subeq{
\frac{1}{\sqrt{-g_3}}\partial_\nu\left(\sqrt{-g_3}g^{\mu\nu}\partial_\mu\sigma\right)&=-\sqrt{2}\, e^{-2\sqrt{2}\sigma-2n\phi}\left(\frac{q}{2\pi R_0}\right)^2\, ,\\
\frac{1}{\sqrt{-g_3}}\partial_\nu\left(\sqrt{-g_3}g^{\mu\nu}\partial_\mu\phi\right)&=-n\, e^{-2\sqrt{2}\sigma-2n\phi}\left(\frac{q}{2\pi R_0}\right)^2\, .
} 
We consider solutions in which the fields run with one of the coordinates $x^3$ (which with hindsight we denote by $r\equiv x^3$).
We focus on a particular 3d  axion-saxion ansatz 
\begin{align}
    s(r) = s_0 - \frac{q}{2\pi} \log\frac{r}{r_0} \quad , \quad    a(r) = a_0 ~.
    \label{sol-s}
\end{align}
for which the radion can be solved as
\begin{equation} \label{eq:4d-sigma}
    \sqrt{2}\sigma= \frac{2}{n} (\phi - \phi_0) + 2\log{\frac{r}{R_0}} \, =\, \frac{2}{n^2} \log\left( 1 - \frac{q}{2\pi s_0} \log\frac{r}{r_0} \right) + 2\log{\frac{r}{R_0}} \, .
\end{equation}
This, together with \eqref{sol-s}, provides the scalar profiles solving the dynamical tadpole. The motivation for this particular solution is that it preserves 1/2 supersymmetry, as we discuss in the next section in the context of its relation with the 4d string solutions in \cite{Lanza:2021qsu}.

Note that as $r\to 0$, the radion blows up as $\sigma\to-\infty$, implying that the $\IS^1$ shrinks to zero size, and the metric becomes singular. As one hits this singularity, the saxion goes to infinity, so we face a wall at which the scalars run off to infinity in moduli space. According to our arguments, it must correspond to a cobordism wall of nothing, capping off spacetime so that the $r<0$ region is absent; hence the suggestive notation to regard this coordinate as a radial one, an interpretation which will become  more clear in the following section. The finite distance $\Delta$ to the wall  can be shown to obey the scaling $\Delta^{-2}\sim {\cal T}$ introduced in \cite{Buratti:2021yia}.

Note that the asymptotic regime near infinity in moduli space $s\gg 1$ corresponds to the regime
\begin{equation} 
\label{eq:regime-validity}
    r \ll r_0 e^{\frac{2\pi}{q} (s_0 - 1)} \, .
\end{equation}
Hence the exploration of the SDC's implications requires zooming into the region close to the wall of nothing.

Let us emphasize that the microscopic structure of the wall of nothing cannot be determined purely in terms of the effective field theory, and should be regarded as provided by its UV completion\footnote{In particular, possible constraints on $q$ could arise from global consistency of the backreaction.}. On the other hand, we can use effective field theory to obtain the scaling relations between different quantities, as in the string theory examples in the previous sections.

\medskip

{\bf The scaling relations}

We can now study the scaling relations between spacetime and moduli space distances, and the SDC tower scale. From the spacetime profiles for $\sigma$ and $\phi$, it is easy to check that the contribution from the radion dominates in the $r\to 0$ limit. The resulting scaling between the moduli space distance $D$ and $r$ is
\begin{equation}
    r \simeq e^{-D} \, ,
\end{equation}
showing again that $D\to \infty$ as $r\to 0$. On the other hand, the spacetime distance $\Delta$ in the same limit gives
\begin{equation}
    d\Delta \simeq \frac{r}{R_0} \left( -\frac{q}{2\pi s_0} \log\left(\frac{r}{r_0}\right) \right)^{\frac{2}{n^2}} \, 
   \simeq \frac{1}{R_0} \left( -\frac{q}{2\pi s_0} \right)^{\frac{2}{n^2}} D^{2/n^2} e^{-2D} dD \, .
\end{equation}
Upon integration one gets an incomplete gamma function that, after keeping the leading order in $D\to\infty$, finally gives
\begin{equation}
    \Delta \sim e^{-2D + \frac{2}{n^2} \log D} \, .
\end{equation}
This is an exponential behaviour up to logarithmic corrections. It would be interesting to relate this to existing results on log corrections to Swampland conjectures (see \cite{Blumenhagen:2019vgj}), but we skip them for now. The resulting relation allows to write the scalings of the SDC tower scale as
\begin{equation}
    M_{\rm SDC} \sim e^{-\lambda D} \sim \Delta^{\frac{\lambda}{2}} \, ,
\end{equation}
that is again a power-like relation with ${\cal O}(1)$ exponents.

Let us turn to computing the scaling of the SDC scale with the scalar curvature $R$. The general expression for $R$ is rather complicated,  but simplifies in the leading order approximation at $r=0$
\begin{equation}
    \log |R| \simeq - 4 \log r \simeq 4\, D \, .
\end{equation}
Hence, the SDC tower mass scales as

\begin{equation}
    M_{\rm SDC} \sim e^{-\lambda D} \sim |R|^{-\frac{1}{4}\lambda} \, .
\end{equation}
Amusingly, we again recover a power-like scaling highly reminiscent of the ADC.

\medskip

{\bf Multi-field generalization}

Let us end this section by mentioning that the above simple model admits a straightforward generalization to several axion-saxion moduli $a^i$, $s^i$. One simply introduces a vector of axion fluxes $q^i$ and generalizes the above running solution to
\beqa
a^i=a_0^i + \frac{\theta}{2\pi}q^i\quad , \quad  s^i(r) = s_0^i - \frac{q^i}{2\pi} \log\frac{r}{r_0}\, .
\eeqa
The corresponding backreaction on $\sigma$ is
\begin{equation}
    \sqrt{2}\,\sigma = -K(r) + K_0 + 2\log\frac{r}{R_0} \, .
\end{equation}
We leave this as an exercise for the reader, since the eventual result is more easily recovered by relating our system to the 4d  string-like solutions in \cite{Lanza:2021qsu}, to which we now turn.

\subsection{Comparison with EFT strings}
\label{sec:eft-strings}

The ansatz  \eqref{sol-s} is motivated by the relation of our setup with the string-like solutions to 4d $\NN=1$ theories discussed in \cite{Lanza:2021qsu}, which we discuss next. This dictionary implies that those results can be regarded as encompassed by our general understanding of cobordism walls of nothing and the SDC.

In a 4d perspective, (\ref{sol-s}) corresponds to a holomorphic profile $z=re^{i\theta}$
\beqa
S\,=\, S_0+\frac{q}{2\pi}\log \frac z{z_0} \, .
\eeqa
The axion flux in (\ref{eq:compactification-ansatz}) implies that there is a monodromy $a\to a+q$ around the origin $z=0$. Hence, the configuration describes a BPS string with $q$ units of axion charge. The solution for the metric can easily be matched with that in \cite{Lanza:2021qsu}. The 4d metric takes the form 
\begin{equation} \label{eq:axionic-metric}
    ds_4 ^2= -dt^2 +dx^2 + e^{2Z}d z d\bar z\, ,
\end{equation}
with the warp factor 
\begin{equation} \label{eq:4d-warping}
    2Z=-K+K_0=\frac{2}{n^2}\log{\frac{s}{s_0}} \, .
\end{equation}
This matches the 3d metric \eqref{eq:axionic-metric} by writing
\begin{equation}\label{3dmetric}
    ds_{3}^{2} = e^{\sqrt{2}\,\sigma} \left( -dt^2 + dx^2 \right) + e^{2Z + \sqrt{2}\,\sigma}dr^2 \, .
\end{equation}
and (\ref{eq:4d-sigma}) ensures the matching of the $S^1$ radion with the 4d angular coordinate range.
\begin{equation} \label{eq:S1-matching}
    \int_0 ^{2\pi} d\theta e^{\sigma/\sqrt{2}} R_0 = \int_0 ^{2\pi} d\theta e^{Z}r\, .
\end{equation}

Hence, in 4d $\NN=1$ theories there is a clear dictionary between running solutions in $\IS^1$ compactifications with axion fluxes and EFT string solutions. The compactification circle maps to the angle around the string; the axion fluxes map to string charges; the coordinate in which fields run (semi-infinite, due to the wall of nothing) maps to the radial coordinate away from the string; the saxion running due to the axion flux induced dynamical tadpole maps to the string backreaction on the saxion, i.e. the string RG flow; the scalars running off to infinity in moduli space as one hits the wall of nothing map to the scalars running off to infinity in moduli space as one reaches the string core. Note that the fact that the wall of nothing is not describable within the effective theory maps to the criterion for an EFT string, i.e. it is regarded as a UV-given defect providing boundary conditions for the effective field theory fields.

This dictionary allows to extend the interesting conclusions in \cite{Lanza:2021qsu} to our context. For instance, the relation between the string tension and its backreaction on the geometry provides a scaling with the spacetime distance. This is the counterpart of the scaling relations we found in our 3d dynamical cobordism discussion in the previous section.

On another line, the Distant Axionic String Conjecture in \cite{Lanza:2021qsu} proposes that every infinite field distance limit of a 4d $\NN=1$ effective theory consistent with quantum gravity can be realized as an RG flow UV endpoint of an EFT string. We can thus map it into the proposal that every infinite field distance limit of a 4d $\NN=1$ effective theory consistent with quantum gravity can be realised as the running into a cobordism wall of nothing in some axion fluxed $\IS^1$ compactification to 3d. It is thus natural to extend this idea to a general conjecture

\smallskip

{\bf Cobordism Distance Conjecture:} {\em Every infinite field distance limit of a effective theory consistent with quantum gravity can be realized as the running into a cobordism wall of nothing in (possibly a suitable compactification of) the theory. }

\smallskip

The examples in the previous sections provide additional evidence for this general form of the conjecture, beyond the above 4d $\NN=1$ context.

\section{4d $\NN=1$ theories with flux-induced superpotentials}
\label{sec:potentials}

In the previous section we discussed cobordism walls in compactifications of 4d $\NN=1$ theories on $\IS^1$ with axion fluxes. Actually, it is also possible to study running solutions and walls in these theories without any compactification. This requires additional ingredients to introduce the dynamical tadpoles triggering the running. Happily, there is a ubiquitous mechanism, via the introduction of non-trivial superpotentials, such as those arising in flux compactifications. We discuss those vacua and their corresponding walls in this section. The discussion largely uses the solutions constructed in \cite{Bandos:2018gjp}, whose notation we largely follow.

Let us consider a theory with a single axion-saxion complex modulus $\Phi=a+iv$. The 4d effective action, in Planck units, is
\beqa
S=-\int d^4x\, \sqrt{-g} \,\Big[\, \frac 12 R + \frac {|\partial \Phi|^2}{4({\rm Im\,} \Phi)^2}\,+\, V(\Phi,{\ov \Phi})\,\Big]
\label{kinetic-dw-4d}
\eeqa
with the $\NN=1$ scalar potential
\ea{ V(\Phi,{\ov \Phi})\, =\, e^{K}\,(\, K^{\Phi{\ov \Phi}}\, |D_\Phi W|^2\, -3|W|^2\,) \, .
}
We focus on theories of the kind considered in \cite{Bandos:2018gjp}, where the superpotential is induced from a set of fluxes $m^I, e_I$, with $I=0,1$, and is given by
\ea{W\,=\, e_If^I(\Phi)-m^I{\cal G}_I(\Phi)
}
for $f^I$, ${\cal G}_I$ some holomorphic functions whose detailed structure we do not need to specify.

In general, these fluxes induce a dynamical tadpole for $\Phi$, unless it happens to sit at the minimum of the potential. The results in \cite{Bandos:2018gjp} allow to build 1/2 BPS running solutions depending on one space coordinate $y$ with
\ea{ds^2\,=\, e^{2Z(y)}dx_\mu dx^\mu\,+\, dy^2 \, .
\label{eq:DW-ansatz}
}

For the profile of the scalar, the solution has constant axion $a$, but varying saxion. Defining the `central charge' ${\cal Z}=e^{{\cal K}/2}\, W$ and ${\cal Z}_*$ its value at the minimum of the potential (and similarly for other quantities), the profile for the scalar $v$ is
\ea{v(y)\,=\, v_*\, \coth^2 \big(\frac 12 |{\cal Z}_*|\, y\,\big) \, .
}
Note that in \cite{Bandos:2018gjp} this solution was built as `the left hand side' of an interpolating domain wall solution (more about it later), but we consider it as the full solution in our setup. Note also that we have shifted the origin in $y$ with respect to the choice in \cite{Bandos:2018gjp}.

The backreaction of the scalar profile on the metric is described by
\ea{Z(y)\,= d\,+\, e^{-\frac 12 {\hat{\cal K}}_0} \Big[\,\log(-\sinh (\frac 12 |{\cal Z}_*|\,y))+\log\cosh (\frac 12 |{\cal Z}_*|y)\,\Big] \, ,
\label{metric-dw-4d}
}
where $d$ is just an integration constant and ${\hat{\cal K}}_0$ is an additive constant in the K\"ahler potential.

The solution exhibits a singularity at $y=0$, which (since the metric along $y$ is flat) is at finite distance in spacetime from other points. On the other hand it is easy to see that the scalar $v$ runs off to infinity as we hit the wall, since
\ea{v(y)\,\rightarrow\, 4\,v_*\, |{\cal Z}_*|^{-2}\, y^{-2}\quad {\rm as}\; y\to 0 \, .
\label{limiting-v}
}
We can obtain the scaling of the moduli space distance with the spacetime distance. Using the kinetic term in (\ref{kinetic-dw-4d}),
\eq{D\,=\,-\int \frac{1}{\sqrt{2}v} \frac{dv}{dy}\, dy\,\simeq\,-\sqrt{2} \log y \,\simeq -\sqrt{2} \log\Delta \, .
}
In the last two equalities we have used (\ref{limiting-v}) and (\ref{eq:DW-ansatz}) respectively. We thus get a familiar power-like scaling for the SDC scale
\eq{M_{\rm SDC}\sim \Delta^{\sqrt{2}\lambda} \, .
}
We also recover the ADC-like scaling with the scalar curvature. At leading order in $y\to0$ one finds
\begin{equation}
    \log |R| \simeq -2 \log y \simeq \sqrt{2} D \, ,
\end{equation}
which gives
\begin{equation}
    M_{\rm SDC}\sim |R|^{-\frac{1}{\sqrt{2}}\lambda} \, .
\end{equation}

This all fits very nicely with our picture that the solution is describing a cobordism wall of nothing, and that the solution for $y>0$ is unphysical and not realized. This provides an effective theory description of the cobordism defects for general 4d $\NN=1$ theories, in a dynamical framework. It would be interesting to find explicit microscopic realizations of this setup.

\medskip

Let us conclude this section by mentioning that it is possible to patch together several solutions of the above kind and build  cobordism domain walls interpolating between different flux vacua. In particular in \cite{Bandos:2018gjp} the solution provided `the left hand side' of one such interpolating domain wall solution whose `right hand side' was glued before reaching (in our choice of origin) $y=0$, hence before encountering the wall of nothing. The particular solution on the right hand side was chosen to sit at the minimum of the corresponding potential, for which there is no tadpole and thus the functions $D$ and $v$ are simply set to constants, fixed to guarantee continuity. Consequently, the solutions remain at finite distance in moduli space, in agreement with our picture for interpolating domain walls. In some sense, the flux changing membrane is absorbing the tadpole, thus avoiding the appearance of the wall of nothing. We refer the reader to \cite{Bandos:2018gjp} (see also \cite{Lanza:2020qmt}) for a detailed discussion.

\section{Walls in 10d non-supersymmetric strings}
\label{sec:nonsusy}

The above examples all correspond to supersymmetric solutions, and even the resulting running solutions preserve some supersymmetry. This is appropriate to establish our key results, but we would like to illustrate that they are not restricted to supersymmetric setups. In order to illustrate that these ideas can apply more generally, and can serve as useful tools for the study of non-supersymmetric theories, we present a quick discussion of the 10d non-supersymmetric $USp(32)$ theory \cite{Sugimoto:1999tx}, building on the solution constructed in \cite{Dudas:2000ff} and revised in \cite{Buratti:2021yia}\footnote{For other references related to dynamical tadpoles in non-supersymmetric theories, see \cite{Blumenhagen:2000dc,Dudas:2002dg,Dudas:2004nd,Mourad:2016xbk,Basile:2018irz,Antonelli:2019nar,Basile:2020xwi}.}.

The 10d (Einstein frame) action reads
\beqa
S_E\,=\, \frac{1}{2\kappa^2}\int d^{10}x \sqrt{-G}\,[\, R-\frac{1}{2}(\partial\phi)^2\,] \,-\, T_9^E\int d^{10}x \sqrt{-G}\, 64\, e^{\frac{3\phi}2}\, ,
\label{action-sugimoto}
\eeqa
where $T_9^E$ is the (anti)D9-brane tension. The theory has a dynamical dilaton tadpole ${\cal T}\sim T_9^E g_s^{3/2}$, and does not admit maximally symmetric solutions. The running solution in \cite{Dudas:2000ff} preserves 9d Poincar\'e invariance, and reads
\beqa
\phi &=& \frac 34 \alpha_E y^2\,+\,\frac 23\log|\sqrt{\alpha_E}y|\,+\,\phi_0 \ ,\nonumber\\
ds_E^{\,2}&=& |\sqrt{\alpha_E}y|^{\frac 19}\, e^{-\frac{\alpha_E y^2}8} \eta_{\mu\nu} dx^\mu dx^\nu +  |\sqrt{\alpha_E}y|^{-1} e^{-\frac{3\phi_0}2} e^{-\frac{9\alpha_E y^2}{8}}\, dy^2\, ,
\label{solution-dm}
\eeqa
where $\alpha_E= 64 k^2T_9$. There are two singularities, at $y=0$ and $y\to\infty$, which despite appearances are located at  finite spacetime distance, satisfying the scaling $\Delta^{-2}\sim {\cal T}$ introduced in \cite{Buratti:2021yia}. In this case, there is no known microscopic description for the underlying cobordism defect, but we can still consider the effective theory solution to study the theory as we hit the walls. 

We consider the two singularities at $y=0,\infty$, and look at the behaviour of the solution near them. The distance from a generic point $y$ to the singularites is given by the integral \cite{Buratti:2021yia}
\eq{\Delta\,\sim\, \int  |\sqrt{\alpha_E}y|^{-\frac 12} e^{-\frac{3\phi_0}4} e^{-\frac{9\alpha_E y^2}{16}}\, dy\ ,}
on the intervals $[y,0]$ when $y\to 0$, and $[y,\infty]$ when $y\to\infty$. They give (lower and upper) incomplete gamma functions
\eq{\Delta_0\sim\gamma\left(\frac{1}{4},\frac{9\alpha_E y^2}{16}\right)\qquad\text{and}\qquad\Delta_\infty\sim\Gamma\left(\frac{1}{4},\frac{9\alpha_E y^2}{16}\right)\ .}
By expanding at leading order as $y\rightarrow0$ and $y\rightarrow\infty$, one gets
\eq{\Delta_0\sim y^\frac{1}{2}\qquad\text{and}\qquad\Delta_\infty\sim y^{-\frac{3}{2}}e^{-\frac{9\alpha_E y^2}{16}}\ .}

The moduli space distance is $\phi=\sqrt{2}D$. Its leading behavior is $D\simeq-\frac{\sqrt{2}}{3}\ln y$ as $y\rightarrow 0^+$ and $D\simeq \frac{3\alpha_E}{4\sqrt{2}} y^2$ as $y\rightarrow\infty$. This leads to the scaling relations
\beqa
y\rightarrow0^+&:& \;\Delta_0\sim e^{-\frac{3}{2\sqrt{2}}D}\, , 
    \nonumber\\
 y\rightarrow\infty&: &\;\Delta_\infty\sim D^{-\frac{3}{4}}e^{-\frac{3}{2\sqrt{2}}D}\sim e^{-\frac{3}{2\sqrt{2}}D-\frac{3}{4}\ln D}\, .
\eeqa
In both cases we have the moduli space distance running off to infinity as we approach the wall. This is in agreement with their interpretation as cobordism walls of nothing\footnote{The interpretation of the $y\to 0$ singularity as a wall of nothing was deemed unconventional, since it would arise at weak coupling. It is interesting that we get additional support for this interpretation from the moduli space distance behaviour.}. Moreover, we recover the a familiar power-like scaling of the SDC mass scale with the same numerical factors in both cases
\beqa
M_{\rm SDC}\sim e^{-\lambda D}\sim \Delta^{\frac{2\sqrt{2}}{3}\lambda} \, .
\eeqa

It is interesting to see that one can also recover a standard power-like scaling for both singularities if the moduli space distance $D$ is compared with the spacetime curvature scalar $R$. The latter reads
\begin{equation}
    |R| = \sqrt{\alpha_E}\, e^{\frac{3\phi_0}{2}} \left( \frac{2}{9} y^{-1} + \frac{7}{2} \alpha_E y + \frac{9}{8} \alpha_E^2 y^3 \right) e^{\frac{9\alpha_E}{8} y^2} \,.
\end{equation}
Let us start with the $y\to 0$ singularity. We can approximate the logarithm of the scalar curvature as
\begin{equation}
    \log |R| \simeq - \log y \simeq \frac{3}{\sqrt{2}} D \, .
\end{equation}
This allows to rewrite the SDC scaling in the form of the ADC-like scaling
\begin{equation}
    M_{\rm SDC} \sim e^{-\lambda \Delta} \sim |R|^{-\frac{\sqrt{2}}{3}\lambda} \, .
\end{equation}

Let us now turn to the $y\to\infty$ limit. In this case the logarithm of the scalar curvature is approximated to
\begin{equation}
    \log |R| \simeq \frac{9\alpha_E}{8} y^2 \simeq \frac{3}{\sqrt{2}} D\, ,
\end{equation}
thus recovering the same behavior as for the other singularity.

As announced, we find a nice power-like scaling, reminiscent as usual of the ADC relations. It is amusing that the precise coefficient arises in both the strong and weak coupling singularities, which may hint towards some universality or duality relation in this non-supersymmetric 10d model.

\section{Final remarks}
\label{sec:conclusions}

In this work we have considered running solutions solving the equations of motion of theories with tadpoles for dynamical fields. These configurations were shown to lead to cobordism walls of nothing at finite distance in spacetime \cite{Buratti:2021yia}, in a dynamical realization of the Cobordism Conjecture. We have also discussed interpolating domain walls across which we change to a different (but cobordant) theory/vacuum. We have shown that the key criterion distinguishing both kinds of walls is related to distance in field space: walls of nothing are characterized by the scalars attaining infinite distance in moduli space, while interpolating domain walls remain at finite distance in moduli space. 

Hence, cobordism walls of nothing provide excellent probes of the structure of the effective theory near infinite distance points, and in particular the Swampland Distance Conjectures. This viewpoint encompasses and generalizes that advocated for EFT strings in 4d $\NN=1$ theories in \cite{Lanza:2021qsu}. We have found interesting new general scaling relations linking, for running solutions, the moduli space distance and the SDC tower mass scale to geometric spacetime quantities, such as the distance to the wall or the scalar curvature. The latter takes a form tantalizingly reminiscent of the Anti de Sitter Distance Conjecture (ADC), suggesting it may relate to the generalized distance in \cite{Lust:2019zwm}.

We have illustrated the key ideas in several large classes of string models, most often in supersymmetric setups (yet with nontrivial scalar potentials to produce the dynamical tadpole triggering the running); however, we emphasize that we expect similar behaviours in non-supersymmetric theories, as we have shown explicitly for the 10d non-susy $USp(32)$ theory.

There are several interesting open question that we leave for future work:

$\bullet$ We have mainly focused on space-dependent running solutions. It is clearly interesting to consider time-dependent solutions, extending existing results in the literature \cite{Dudas:2000ff,Blumenhagen:2000dc,Dudas:2002dg,Dudas:2004nd,Mourad:2016xbk,Basile:2018irz,Antonelli:2019nar,Basile:2020xwi}, and exploit them in applications, in particular with an eye on possible implications for inflationary models or quintessence.

$\bullet$ A particular class of time-dependent solutions are dynamical bubbles. In particular, a tantalizing observation is that in the original bubble of nothing in \cite{Witten:1981gj}, the 4d radion modulus goes to zero size (which lies at infinite distance in moduli space of the $\IS^1$ compactification) as one hits the wall. Although the setup is seemingly unrelated, it would be interesting to understand universal features of bubbles of nothing along the lines considered in our work.

$\bullet$ The appearance of ADC-like scaling relations in our running solutions possibly signals an underlying improved understanding of infinite distance limits in dynamical (rather than adiabatic) configurations. For instance, as shown in \cite{Buratti:2018xjt}, the $r\to\infty$ limit in the Klebanov-Strassler  solution \cite{Klebanov:2000hb} avoids the appearance of a tower of states becoming massless exponentially with the distance. This was related to having a non-geodesic trajectory in moduli space (see \cite{Calderon-Infante:2020dhm} for a general discussion about non-geodesics and the SDC). However, as dictated by the lack of separation of scales in this model, an ADC-like scaling is yet respected as the scalar curvature goes to zero in this limit. This could point to a more universal way of writing the SDC in dynamical configurations. 

$\bullet$ In all the examples we find precise numbers relating the parameter in the SDC $\lambda$ to the power in the ADC-like scaling. It would certainly be interesting to find a pattern in these values and possibly relate them to properties of the infinite distance limits along the lines of \cite{Grimm:2018ohb,Corvilain:2018lgw,Grimm:2019ixq}. On a similar line of thought, it has been argued that in supersymmetric cases the ADC's scaling parameter should be $1/2$ \cite{Lust:2019zwm}, assuming this applies to our setup, it would be interesting to extract the SDC's parameter $\lambda$ from our supersymmetric examples with an ADC-like scaling. It would be remarkable that they match the existing proposals for the value of $\lambda$.

$\bullet$ The ADC-like scaling may also signal some potential interplay with the Gravitino Distance Conjecture \cite{Cribiori:2021gbf,Castellano:2021yye}. One expects to find a power relation between the mass of the gravitino and the scalar curvature of the solution, it would be certainly interesting to test this and to look for some pattern in the corresponding powers. 

$\bullet$ The trajectory in moduli space in spacetime-dependent solutions has a strong presence in the study of black holes, in particular attractor equations and flows. The attempts to relate them to the SDC (see e.g. \cite{Bonnefoy:2019nzv}) can have an interesting interplay with our general framework.

$\bullet$ We certainly expect interesting new applications of our results to the study of non-supersymmetric strings, and to supersymmetry-breaking configurations in string theory.

We hope to report on these problems in the near future.

%\newpage
%
\section*{Acknowledgments}
We are pleased to thank Ivano Basile, Inaki Garc\'ia-Etxebarria, Luis Ib\'anez, Fernando Marchesano, Miguel Montero, Pablo Soler and Irene Valenzuela for useful discussions. This work is is supported by the Spanish Research Agency (Agencia Espa\~nola de Investigaci\'on) through the grants IFT Centro de Excelencia Severo Ochoa SEV-2016-0597, the grant GC2018-095976-B-C21 from MCIU/AEI/FEDER, UE. The work by J.C. is supported by the FPU grant no. FPU17/04181 from Spanish Ministry of Education.

\newpage

\appendix

\section{Holographic examples}
\label{app:holographic}

In \cite{Buratti:2021yia} it was shown that Dynamical Cobordism underlies the structure of the gravity dual of the $SU(N)\times SU(N+M)$ conifold theory, namely fractional brane deformation of AdS$_5\times T^{1,1}$. This in fact explains the appearance of a singularity at finite radial distance \cite{Klebanov:2000nc} and its smoothing out into a configuration capping of the warped throat \cite{Klebanov:2000hb}, as a cobordism wall of nothing. In this appendix we provide some examples of other warped throat configurations which illustrate the appearance of other cobordism walls of nothing, and cobordism domain walls interpolating between theories corresponding to compactification on horizons of different topology. The discussion is strongly inspired by the constructions in \cite{Franco:2005fd} (see also \cite{GarciaEtxebarria:2006aq}).

\subsection{Domain wall to a new vacuum}
\label{sec:wall-to-ads}

As a first example we consider a configuration in which a running of the conifold theory hits a wall (given by the tip of a KS throat) interpolating to an AdS$_5\times \IS^5$ vacuum. The latter is the maximally symmetric solution of a theory at the bottom of its potential, i.e. with no dynamical tadpole. We carry out the discussion in terms of the dual field theory, which translates easily into the just explained gravity picture. The dilaton is constant in the whole configuration, so we skip factors of $g_s$.

Consider the conifold theory with $SU(N) \times SU(N+M)$ at some scale, i.e. at some position $r$ there are $N$ units of 5-form flux and $M$ units of 3-form flux. The Klebanov-Tseytlin solution \cite{Klebanov:2000nc} gives a running for the effective flux
\beqa
N(r)= N +M^2 \log (r)\, ,
\eeqa
 and we get a singu at a value $r_0$ defined by 
 \beqa
 N+M^2 \log r_0 = 0\quad \Rightarrow \quad r_0 =  e^{-N/M^2} \, .
 \eeqa
Naively,  the singularity would seem to be smoothed out into a purely geometric background with a finite size $\IS^3$. Indeed, this is the full story if $N$ is multiple of $M$, namely $N=KM$: in the field theory, the $SU(KM) \times SU(KM+M)$ theory suffers a cascade of $K$ Seiberg dualities in which $K$ decreases by one unit in each step. Morally, the cascade ends when the effective $K=0$ and then we just have a pure $SU(M)$ SYM, which confines and develops a mass gap. This is the end of the RG flow, with no more running, hence the spacetime in capped off in the IR region of the dual throat.

\begin{figure}[htb]
\begin{center}
\includegraphics[scale=.55]{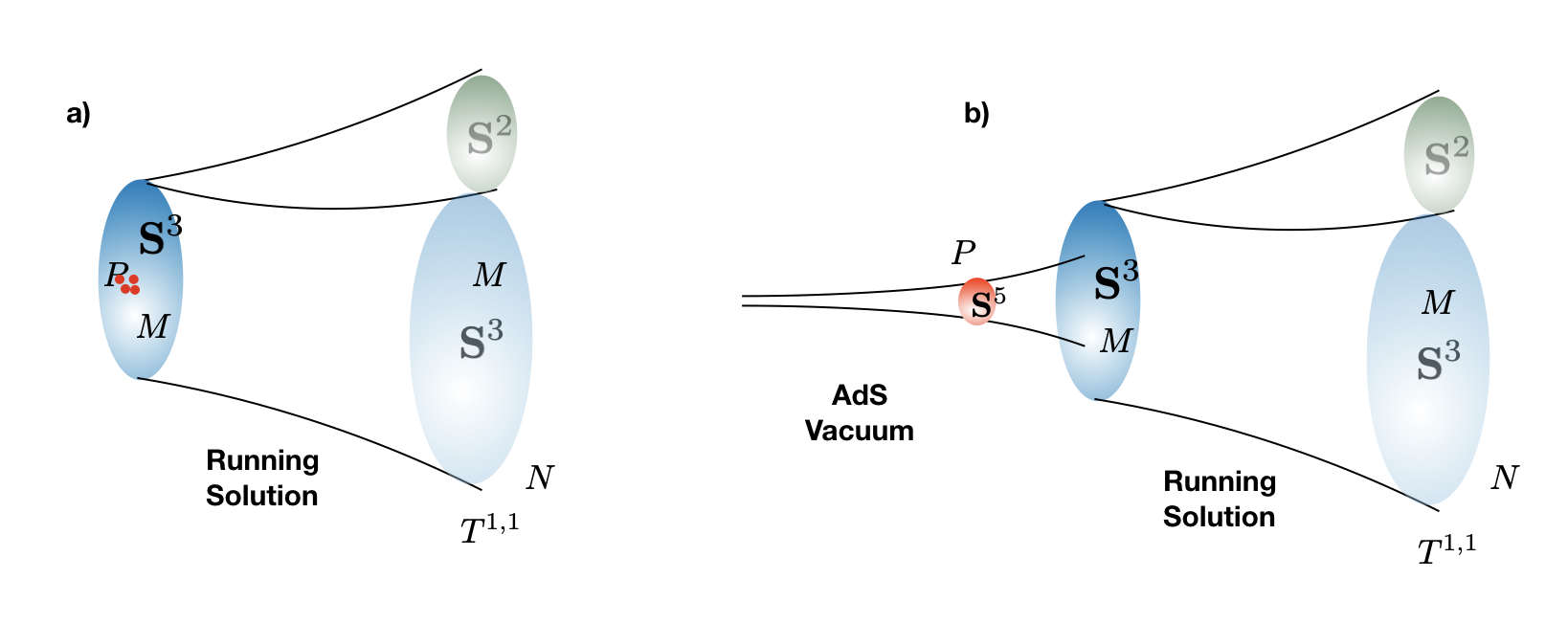}
\caption{\small Domain wall interpolating between the conifold theory with fractional branes, and an AdS vacuum. Figure a) shows a heuristic intermediate step of a KS solution with a number $P$ of left-over probe D3-branes. If $P$ is large, the appropriate description requires including the backreaction of the D3-branes, which lead to a further AdS throat, to the left of the picture in Figure b). Hence the running of the dynamical tadpole in the right hand side ends in a domain wall separating it from an AdS vacuum.}
\label{fig:brave-new-vacuum}
\end{center}
\end{figure}

However, as also noticed in \cite{Klebanov:2000hb}, the story is slightly different if $N=KM+P$. After the $K$ steps in the duality cascade, one is left over with an $SU(P)$ gauge theory with three complex scalar degrees of freedom parametrizing a deformed mesonic moduli space corresponding to (the symmetrization of $P$ copies of) the deformed conifold. This gauge theory flows to $\NN=4$ $SU(P)$ SYM in the infrared, which is a conformal theory. In the parameter range $1\ll P\ll M\ll N$, the whole configuration admits a weakly coupled supergravity dual given by a KS throat at which infrared region we have a finite size $\IS^3$, at which $P$ D3-branes (which we take coincident) would be located; however, since $P$ is large, they backreact and carve out a further AdS$_5\times \IS^5$ with $P$ units of RR 5-form flux, which continues the radial direction beyond the KS throat endpoint region. Hence, this region acts as an interpolating domain wall between two different (but cobordant) theories, namely the conifold throat (with a dynamical tadpole from the fractional brane charge), and the AdS$_5\times \IS^5$ vacuum (with no tadpole, and preserving maximal symmetry). The picture is summarized in Figure \ref{fig:brave-new-vacuum}

%%%%%%%%%%%
%%%%%%%%%%%

\subsection{Domain wall to a new running solution}

Running can lead to an interpolating domain wall, across which we find not a vacuum, but a different running solution (subsequently hitting a wall of nothing, other interpolating domain walls, or just some AdS vacuum). We now illustrate this idea with an example of a running solution A hitting a domain wall interpolating to a second running solution B, which subsequently hits a wall of nothing. The example is based on the multi-flux throat construction in \cite{Franco:2005fd} (whose dimer picture is given in \cite{Franco:2005zu}). It is easy to devise other generalizations displaying the different behaviours mentioned above.

\begin{figure}[htb]
\begin{center}
\includegraphics[scale=.4]{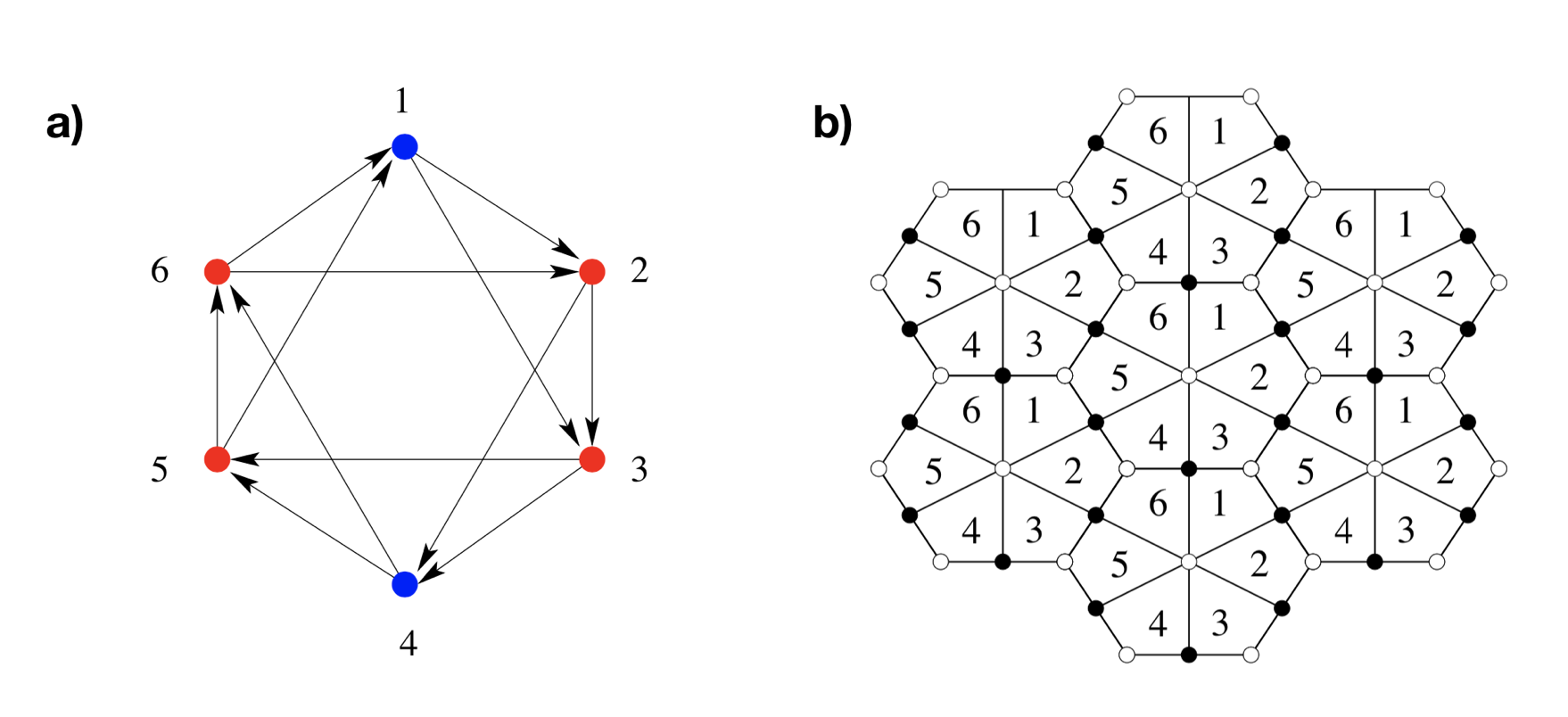}
\caption{\small The quiver and dimer diagrams describing the gauge theory on D3-branes at the tip of the complex cone over $dP_3$.}
\label{fig:dp3-quiver-dimer}
\end{center}
\end{figure}

Consider the system of D3-branes at the singularity given by the complex cone over $dP_3$. The gauge theory is described by the quiver and dimer diagrams\footnote{For references, see \cite{Hanany:2005ve,Franco:2005rj,GarciaEtxebarria:2006aq,Kennaway:2007tq}.} in Figure \ref{fig:dp3-quiver-dimer}.

%%%%%%%%%%%
%%%%%%%%%%%

\begin{figure}[htb]
\begin{center}
\includegraphics[scale=.4]{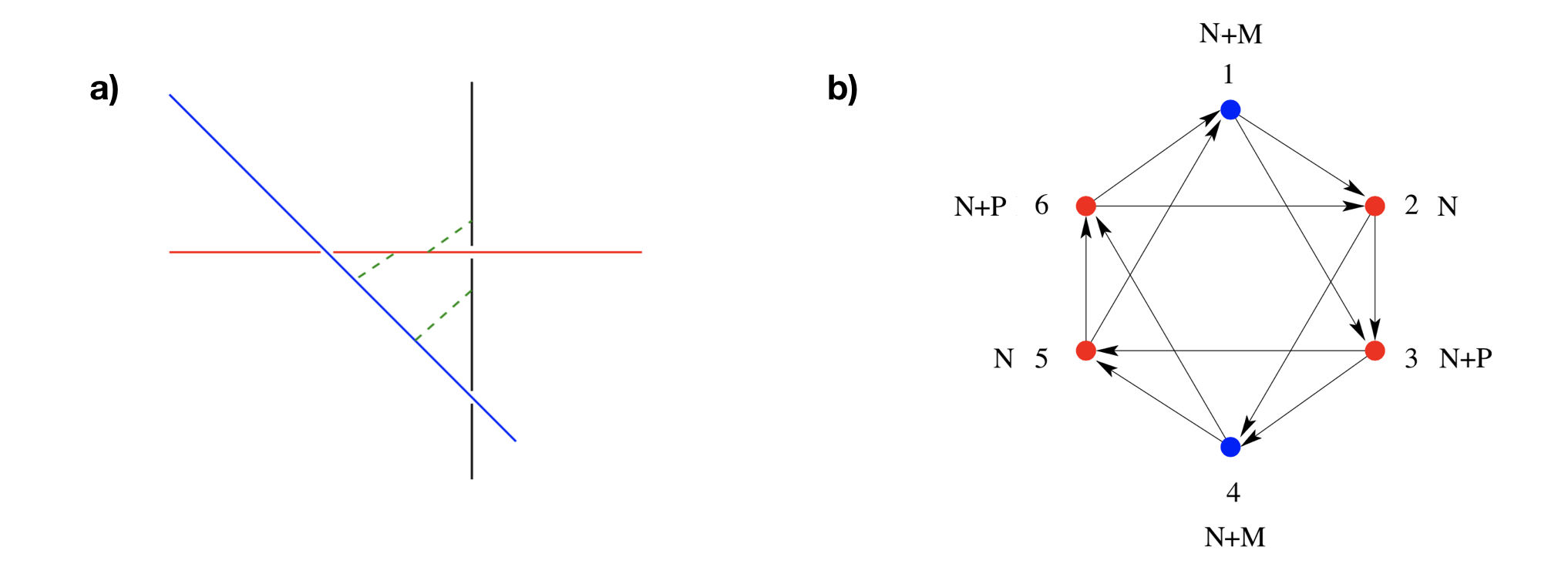}
\caption{\small a) Web diagram of the complex cone over $dP_3$ splitting into three sub-webs. b) Rank assignment (fractional branes) that trigger those complex deformations.}
\label{fig:deformed-dp3}
\end{center}
\end{figure}

We can add fractional branes, i.e. rank assignments compatible with cancellation of non-abelian anomalies. There are several choices, corresponding to different fluxes on the 3-cycles in the dual gravitational theory. Some of them correspond to 3-cycles which can be grown out of the singular origin to provide a complex deformation of the CY. These are described as the splitting of the  web diagram into sub-webs in equilibrium, see \cite{Franco:2005zu}. In particular we focus on the complex deformation of complex cone over $dP_3$ to a conifold, see the web diagram in Figure \ref{fig:deformed-dp3}.

%%%%%%%%%%%
%%%%%%%%%%%

There are two kinds of fractional branes, associated to $M$ and $P$. In the gravity dual, these correspond to RR 3-form fluxes on 3-cycles (obtained by an $\IS^1$ fibration over a 2-cycle on $dP_3$), and there are NSNS 3-form fluxes in the dual 3-cycles. These are non-compact, namely they span a 2-cycle (dual to the earlier 2-cycle in $dP_3$) and the radial direction. For more details about the quantitative formulas of this kind of solution, see Section 5 of \cite{Franco:2004jz}.

\begin{figure}[htb]
\begin{center}
\includegraphics[scale=.4]{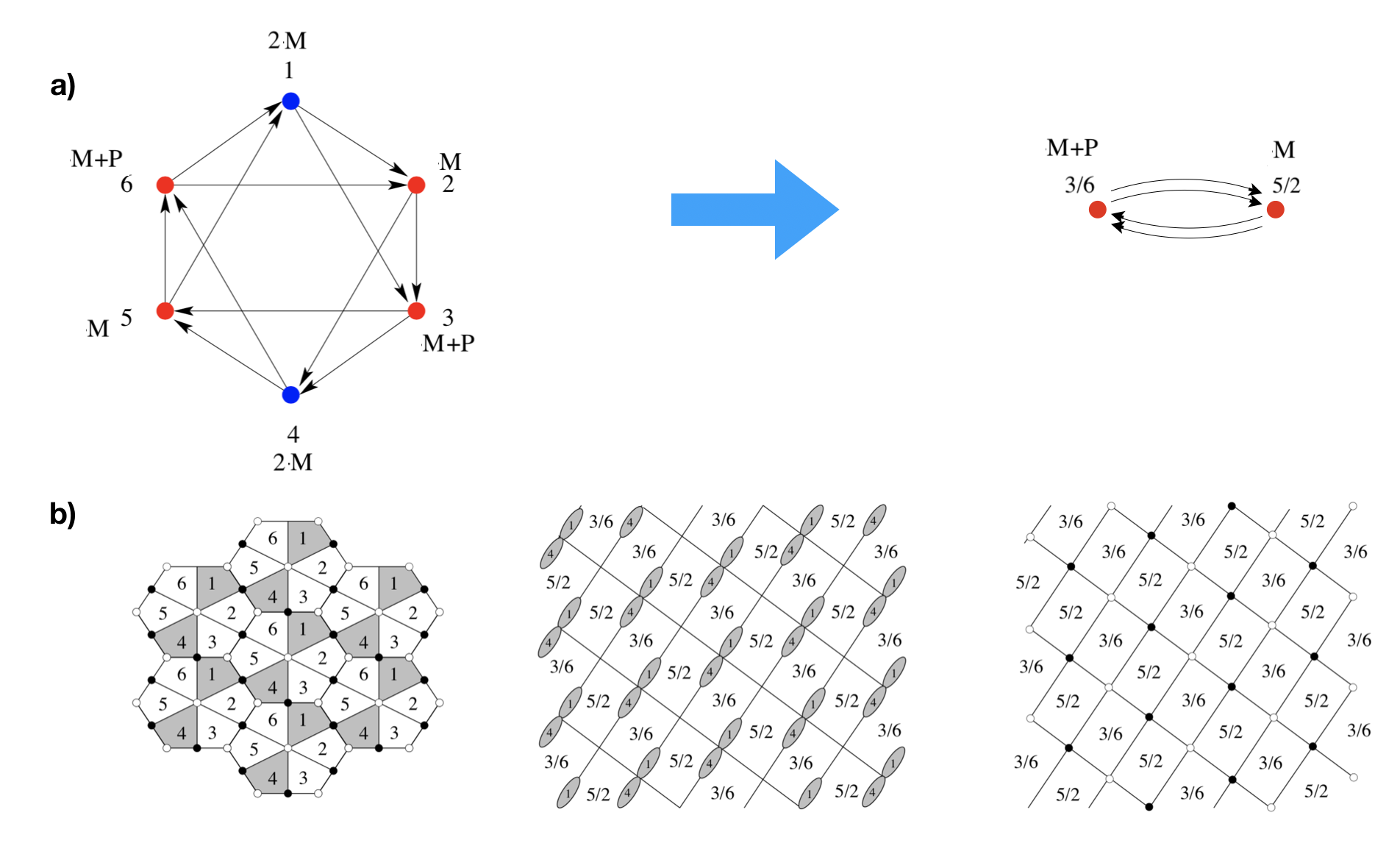}
\caption{\small a) Quiver of the $dP_3$ theory in the last step of the first cascade, which turns into the conifold upon strong dynamics of the nodes 1 and 4. b) Same story in the dimer picture.}
\label{fig:dp3-to-conifold}
\end{center}
\end{figure}

If we focus in the regime\footnote{Note that in  \cite{Franco:2005fd} the regime is the opposite, but both kinds of fractional branes are similar, so the result is the same up to relabeling.} $P\ll M$, then the larger flux $M$ implies a larger corresponding component of the $H_3$ flux, which means a faster running of the corresponding 5d NSNS axion. The axion associated to the flux $P$ also runs, but more slowly. In the field theory, the duality cascade is controlled by $M$, so that $N$ is reduced in multiples of $M$ (at leading order in $P/M$). When $N$ is exhausted we are left with a rank assignment as given in Figure \ref{fig:dp3-to-conifold}a. The result of the strong dynamics triggered by $M$ can be worked out in field theory as in \cite{Franco:2005fd} or using dimers as in \cite{Franco:2005zu}. All the info about this last description is in Figure \ref{fig:dp3-to-conifold}b. 

%%%%%%%%%%%
%%%%%%%%%%%

\begin{figure}[htb]
\begin{center}
\includegraphics[scale=.35]{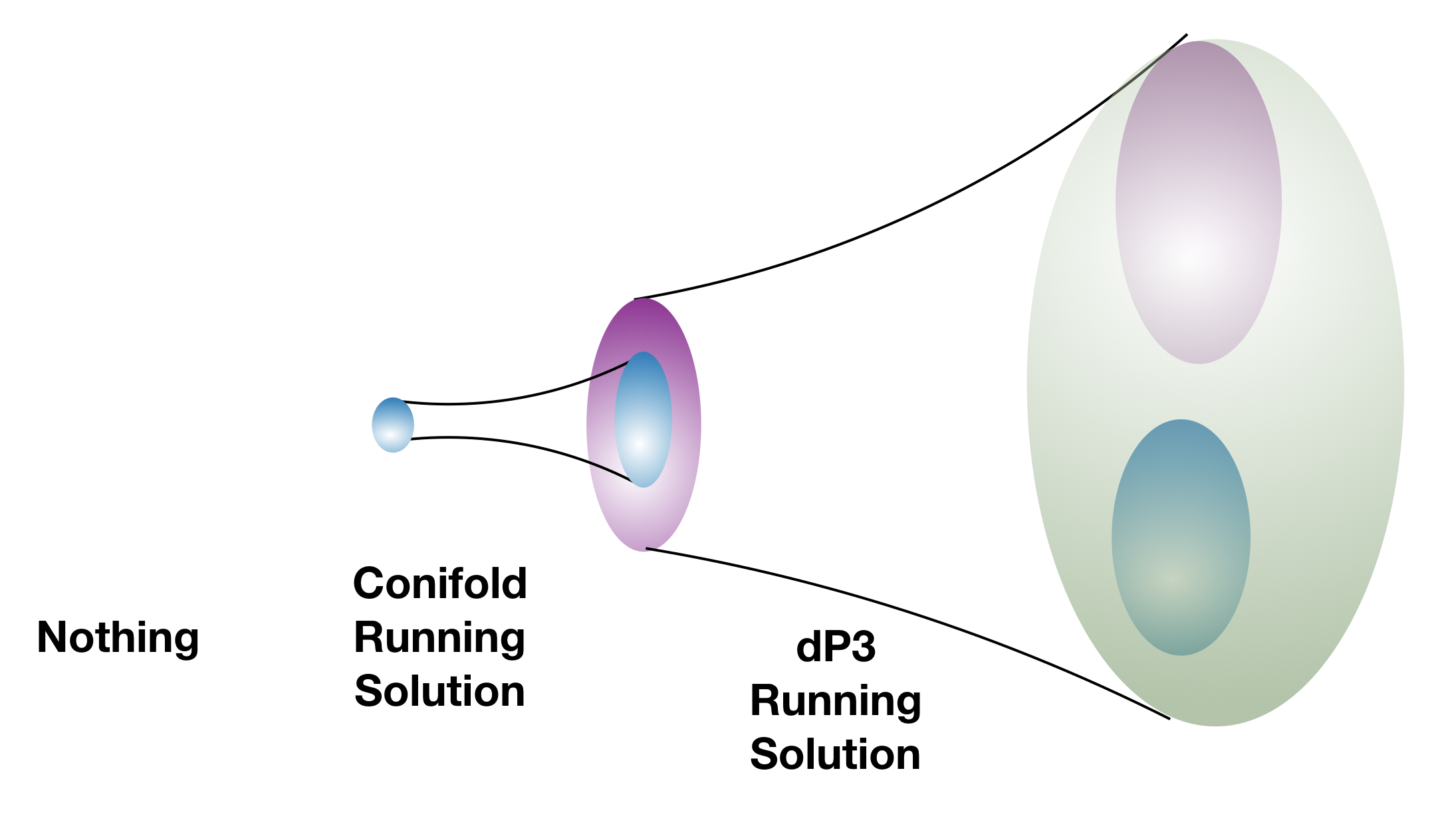}
\caption{\small Domain wall interpolating between the theory on $dP_3$ with $(M+P)$ fractional branes, and a conifold theory with M regular branes and P fractional ones. The running of one of the dynamical tadpoles in the $dP_3$ theory stops at the wall but the other continues running until it reaches the $\IS^3$ at the bottom of the KS throat.}
\label{fig:double-throat}
\end{center}
\end{figure}

The result is a conifold theory with $M$ regular branes and $P$ fractional branes. This is the standard KS story (with just different labels for the branes): $M$ decreases in sets of $P$ until it is exhausted, then the running stops due to strong dynamics. In the gravity dual, we have a KS throat sticking out and spacetime ends on the usual $\IS^3$ (alternatively, if $M$ is not a multiple of $P$, there is a number $P$ of leftover D3-branes, which, if large, can trigger a further AdS throat as in Section \ref{sec:wall-to-ads}. A sketch of the gravity dual picture is shown in Figure \ref{fig:double-throat}.

%%%%%%%%%%%
%%%%%%%%%%%

Note that this kind of domain wall interpolates into two topologically different compactifications. As we cross it, the compactification space changes, and the spectrum of light fields changes (at the massless level, one of the axions ceases to exist). In this sense, it is a cobordism domain wall connecting two different quantum gravity theories \cite{McNamara:2019rup}.

\subsection{Cobordism domain walls to disconnected solutions}

The construction of singularities admitting complicated patterns of complex deformations (or resolutions) can be carried out systematically for toric singularities, using the techniques in \cite{GarciaEtxebarria:2006aq}. This can be used to build sequences of domain walls realizing a plethora of possibilities. For our last class of examples, we consider cobordism domain walls to disconnected theories. 

\begin{figure}[htb]
\begin{center}
\includegraphics[scale=.4]{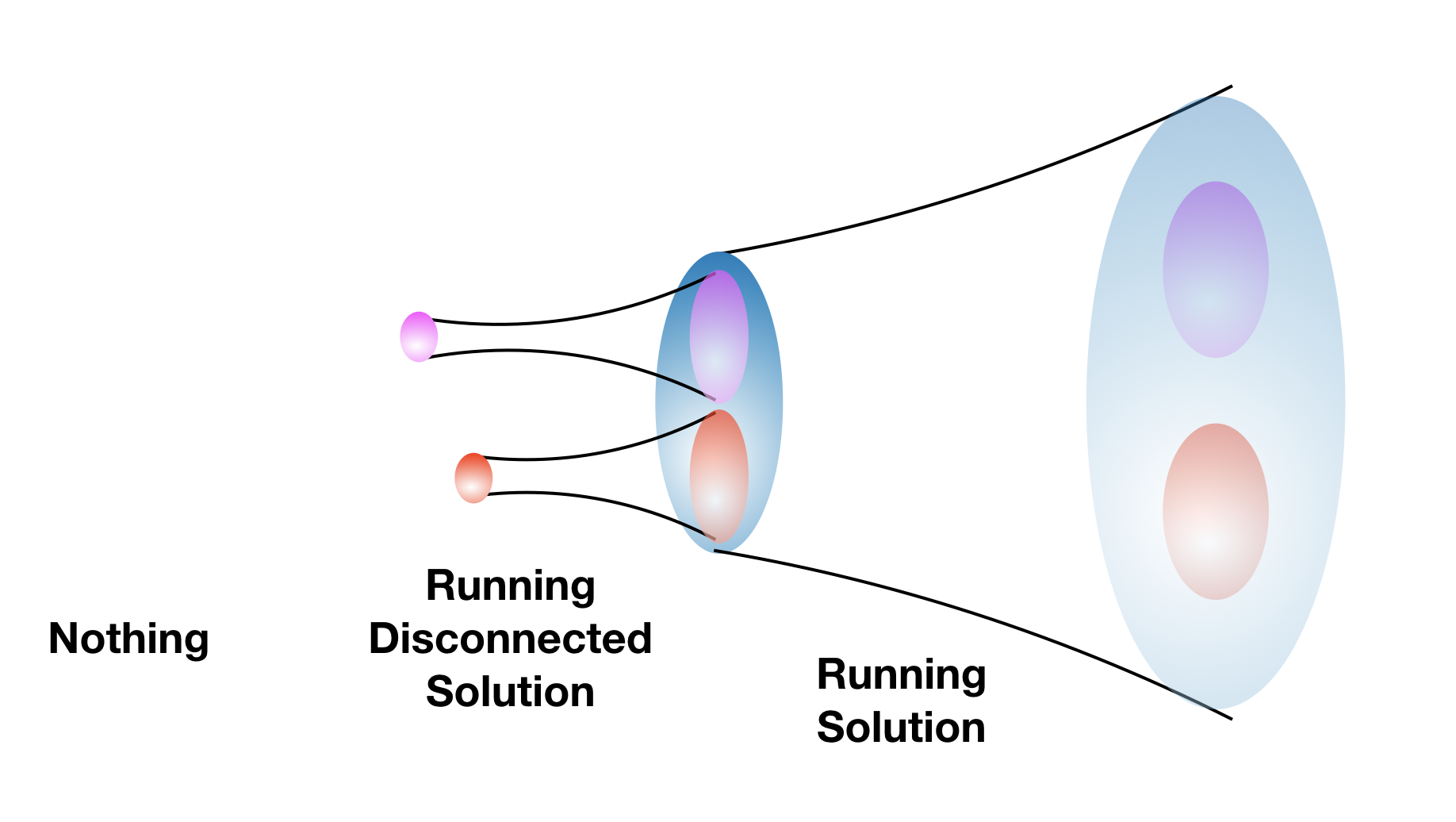}
\caption{\small Picture of a bifid throat. It represents a domain wall implementing a cobordism between one theory and a disconnected set of two quantum gravity theories.}
\label{fig:bifid-throat}
\end{center}
\end{figure}

This has already been realized in the geometry used in \cite{Retolaza:2015sta} to build a bifid throat, i.e. two throats at the bottom of a throat, see Figure \ref{fig:bifid-throat}. These had been proposed in \cite{McAllister:2008hb} as possible hosts of axion monodromy inflation models (see \cite{Silverstein:2008sg,Kaloper:2008fb,Marchesano:2014mla,Hebecker:2014eua,McAllister:2014mpa} for additional references).

%%%%%%%%%%%
%%%%%%%%%%%

\begin{figure}[htb]
\begin{center}
\includegraphics[scale=.4]{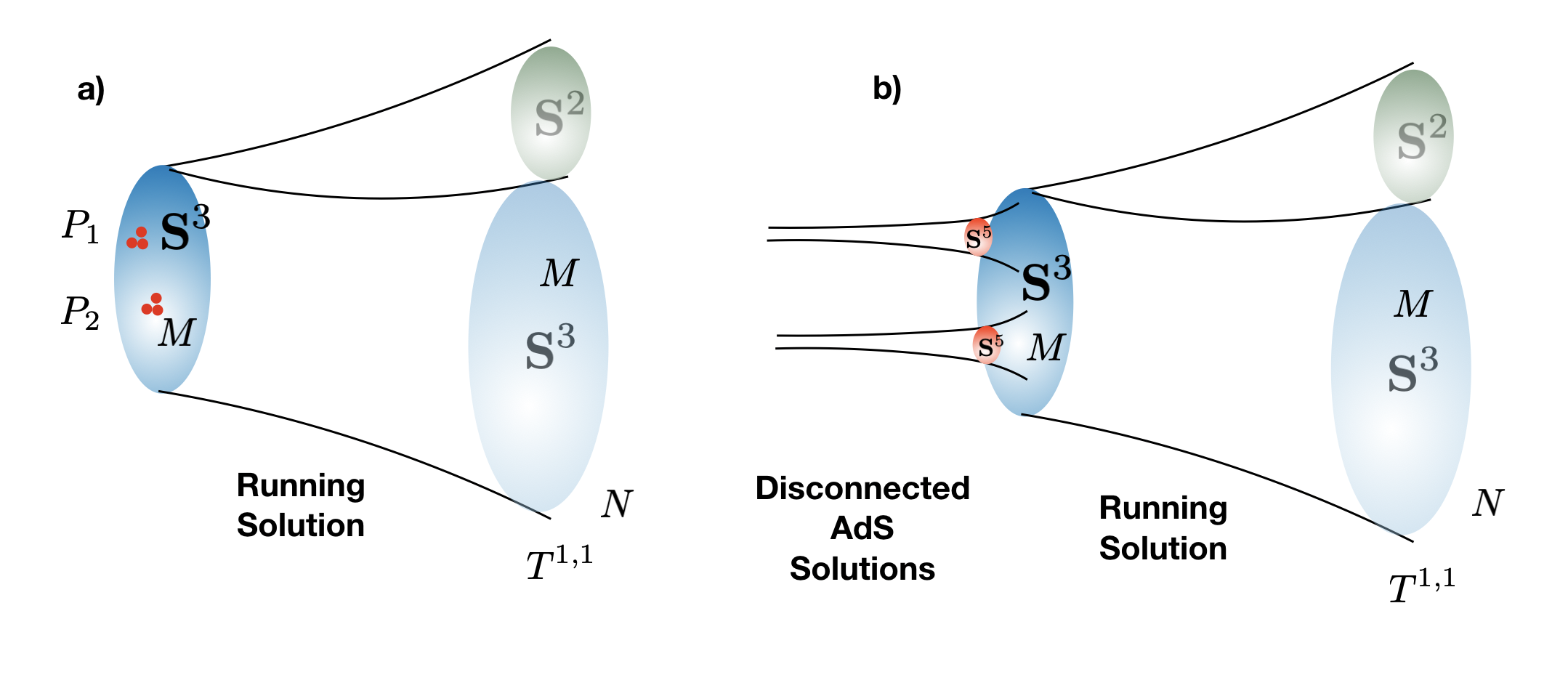}
\caption{\small Picture of a bifid throat with two AdS tongues. It represents a domain wall implementing a cobordism between one theory and a disconnected set of two AdS theories.}
\label{fig:bifid-ads}
\end{center}
\end{figure}

Actually, a far simpler way of getting a running solution with a domain wall to a disconnected set of e.g. vacua is to consider the KS setup in Section \ref{sec:wall-to-ads}, with the $P$ leftover D3-branes split into two stacks $P_1$ and $P_2$ of D3-branes at separated locations on the $\IS^3$ (with $P_1,P_2\gg 1$). This corresponds to turning on a vev $v$ for a Higgsing $SU(P)\to SU(P_1)\times SU(P_2)$ (with $P_1+P_2=P$) with a scale for $v$ much smaller than the scale of confinement $\Lambda$ of the original $SU(KM+P)\times SU(KM+M+P)$ theory. In the gravity dual, we have a running solution in the holographic direction, towards low energies; upon reaching $\Lambda$, we have the $\IS^3$ domain wall, out of which we have one AdS$_5\times S^5$-like vacuum (with flux $P$), until we hit the scale $v$, and the single throat splits into two AdS$_5\times \IS^5$ throats (with fluxes $P_1$, $P_2$). If $v\simeq \Lambda$, the splitting of throats happens in the same regime as the domain wall ending the run of the initial solution. This is depicted in Figure \ref{fig:bifid-ads}.

%%%%%%%%%%%
%%%%%%%%%%%

\bibliographystyle{JHEP}
\bibliography{mybib}

\providecommand{\href}[2]{#2}\begingroup\raggedright\begin{thebibliography}{10}

\bibitem{Vafa:2005ui}
C.~Vafa, {\it {The String landscape and the swampland}},
  \href{http://arxiv.org/abs/hep-th/0509212}{{\tt hep-th/0509212}}.

\bibitem{Brennan:2017rbf}
T.~D. Brennan, F.~Carta, and C.~Vafa, {\it {The String Landscape, the
  Swampland, and the Missing Corner}},  {\em PoS} {\bf TASI2017} (2017) 015,
  [\href{http://arxiv.org/abs/1711.00864}{{\tt arXiv:1711.00864}}].

\bibitem{Palti:2019pca}
E.~Palti, {\it {The Swampland: Introduction and Review}},  {\em Fortsch. Phys.}
  {\bf 67} (2019), no.~6 1900037, [\href{http://arxiv.org/abs/1903.06239}{{\tt
  arXiv:1903.06239}}].

\bibitem{vanBeest:2021lhn}
M.~van Beest, J.~Calder\'on-Infante, D.~Mirfendereski, and I.~Valenzuela, {\it
  {Lectures on the Swampland Program in String Compactifications}},
  \href{http://arxiv.org/abs/2102.01111}{{\tt arXiv:2102.01111}}.

\bibitem{McNamara:2019rup}
J.~McNamara and C.~Vafa, {\it {Cobordism Classes and the Swampland}},
  \href{http://arxiv.org/abs/1909.10355}{{\tt arXiv:1909.10355}}.

\bibitem{Witten:1981gj}
E.~Witten, {\it {Instability of the Kaluza-Klein Vacuum}},  {\em Nucl. Phys. B}
  {\bf 195} (1982) 481--492.

\bibitem{Ooguri:2017njy}
H.~Ooguri and L.~Spodyneiko, {\it {New Kaluza-Klein instantons and the decay of
  AdS vacua}},  {\em Phys. Rev.} {\bf D96} (2017), no.~2 026016,
  [\href{http://arxiv.org/abs/1703.03105}{{\tt arXiv:1703.03105}}].

\bibitem{GarciaEtxebarria:2020xsr}
I.~n. Garc\'\i{}a~Etxebarria, M.~Montero, K.~Sousa, and I.~Valenzuela, {\it
  {Nothing is certain in string compactifications}},  {\em JHEP} {\bf 12}
  (2020) 032, [\href{http://arxiv.org/abs/2005.06494}{{\tt arXiv:2005.06494}}].

\bibitem{Montero:2020icj}
M.~Montero and C.~Vafa, {\it {Cobordism Conjecture, Anomalies, and the String
  Lamppost Principle}},  {\em JHEP} {\bf 01} (2021) 063,
  [\href{http://arxiv.org/abs/2008.11729}{{\tt arXiv:2008.11729}}].

\bibitem{Tarazi:2021duw}
H.-C. Tarazi and C.~Vafa, {\it {On The Finiteness of 6d Supergravity
  Landscape}},  \href{http://arxiv.org/abs/2106.10839}{{\tt arXiv:2106.10839}}.

\bibitem{Buratti:2021yia}
G.~Buratti, M.~Delgado, and A.~M. Uranga, {\it {Dynamical tadpoles, stringy
  cobordism, and the SM from spontaneous compactification}},  {\em JHEP} {\bf
  06} (2021) 170, [\href{http://arxiv.org/abs/2104.02091}{{\tt
  arXiv:2104.02091}}].

\bibitem{Mininno:2020sdb}
A.~Mininno and A.~M. Uranga, {\it {Dynamical Tadpoles and Weak Gravity
  Constraints}},  \href{http://arxiv.org/abs/2011.00051}{{\tt
  arXiv:2011.00051}}.

\bibitem{Dudas:2000ff}
E.~Dudas and J.~Mourad, {\it {Brane solutions in strings with broken
  supersymmetry and dilaton tadpoles}},  {\em Phys. Lett. B} {\bf 486} (2000)
  172--178, [\href{http://arxiv.org/abs/hep-th/0004165}{{\tt hep-th/0004165}}].

\bibitem{Blumenhagen:2000dc}
R.~Blumenhagen and A.~Font, {\it {Dilaton tadpoles, warped geometries and large
  extra dimensions for nonsupersymmetric strings}},  {\em Nucl. Phys. B} {\bf
  599} (2001) 241--254, [\href{http://arxiv.org/abs/hep-th/0011269}{{\tt
  hep-th/0011269}}].

\bibitem{Dudas:2002dg}
E.~Dudas, J.~Mourad, and C.~Timirgaziu, {\it {Time and space dependent
  backgrounds from nonsupersymmetric strings}},  {\em Nucl. Phys. B} {\bf 660}
  (2003) 3--24, [\href{http://arxiv.org/abs/hep-th/0209176}{{\tt
  hep-th/0209176}}].

\bibitem{Dudas:2004nd}
E.~Dudas, G.~Pradisi, M.~Nicolosi, and A.~Sagnotti, {\it {On tadpoles and
  vacuum redefinitions in string theory}},  {\em Nucl. Phys. B} {\bf 708}
  (2005) 3--44, [\href{http://arxiv.org/abs/hep-th/0410101}{{\tt
  hep-th/0410101}}].

\bibitem{Mourad:2016xbk}
J.~Mourad and A.~Sagnotti, {\it {$AdS$ Vacua from Dilaton Tadpoles and Form
  Fluxes}},  {\em Phys. Lett. B} {\bf 768} (2017) 92--96,
  [\href{http://arxiv.org/abs/1612.08566}{{\tt arXiv:1612.08566}}].

\bibitem{Basile:2018irz}
I.~Basile, J.~Mourad, and A.~Sagnotti, {\it {On Classical Stability with Broken
  Supersymmetry}},  {\em JHEP} {\bf 01} (2019) 174,
  [\href{http://arxiv.org/abs/1811.11448}{{\tt arXiv:1811.11448}}].

\bibitem{Antonelli:2019nar}
R.~Antonelli and I.~Basile, {\it {Brane annihilation in non-supersymmetric
  strings}},  {\em JHEP} {\bf 11} (2019) 021,
  [\href{http://arxiv.org/abs/1908.04352}{{\tt arXiv:1908.04352}}].

\bibitem{Basile:2020xwi}
I.~Basile, {\em {On String Vacua without Supersymmetry: brane dynamics, bubbles
  and holography}}.
\newblock PhD thesis, Scuola normale superiore di Pisa, Pisa, Scuola Normale
  Superiore, 2020.
\newblock \href{http://arxiv.org/abs/2010.00628}{{\tt arXiv:2010.00628}}.

\bibitem{Buratti:2018xjt}
G.~Buratti, J.~Calder{\'o}n, and A.~M. Uranga, {\it {Transplanckian axion
  monodromy!?}},  {\em JHEP} {\bf 05} (2019) 176,
  [\href{http://arxiv.org/abs/1812.05016}{{\tt arXiv:1812.05016}}].

\bibitem{Ooguri:2006in}
H.~Ooguri and C.~Vafa, {\it {On the Geometry of the String Landscape and the
  Swampland}},  {\em Nucl. Phys.} {\bf B766} (2007) 21--33,
  [\href{http://arxiv.org/abs/hep-th/0605264}{{\tt hep-th/0605264}}].

\bibitem{Lust:2019zwm}
D.~L{\"u}st, E.~Palti, and C.~Vafa, {\it {AdS and the Swampland}},
  \href{http://arxiv.org/abs/1906.05225}{{\tt arXiv:1906.05225}}.

\bibitem{Lanza:2021qsu}
S.~Lanza, F.~Marchesano, L.~Martucci, and I.~Valenzuela, {\it {The EFT stringy
  viewpoint on large distances}},  \href{http://arxiv.org/abs/2104.05726}{{\tt
  arXiv:2104.05726}}.

\bibitem{Lanza:2020qmt}
S.~Lanza, F.~Marchesano, L.~Martucci, and I.~Valenzuela, {\it {Swampland
  Conjectures for Strings and Membranes}},
  \href{http://arxiv.org/abs/2006.15154}{{\tt arXiv:2006.15154}}.

\bibitem{Polchinski:1995df}
J.~Polchinski and E.~Witten, {\it {Evidence for heterotic - type I string
  duality}},  {\em Nucl. Phys.} {\bf B460} (1996) 525--540,
  [\href{http://arxiv.org/abs/hep-th/9510169}{{\tt hep-th/9510169}}].

\bibitem{Witten:1996mz}
E.~Witten, {\it {Strong coupling expansion of Calabi-Yau compactification}},
  {\em Nucl. Phys. B} {\bf 471} (1996) 135--158,
  [\href{http://arxiv.org/abs/hep-th/9602070}{{\tt hep-th/9602070}}].

\bibitem{Greene:2000yb}
B.~R. Greene, K.~Schalm, and G.~Shiu, {\it {Dynamical topology change in M
  theory}},  {\em J. Math. Phys.} {\bf 42} (2001) 3171--3187,
  [\href{http://arxiv.org/abs/hep-th/0010207}{{\tt hep-th/0010207}}].

\bibitem{Romans:1985tz}
L.~J. Romans, {\it {Massive N=2a Supergravity in Ten-Dimensions}},  {\em Phys.
  Lett. B} {\bf 169} (1986) 374.

\bibitem{Bergshoeff:1995vh}
E.~Bergshoeff, M.~B. Green, G.~Papadopoulos, and P.~K. Townsend, {\it {The IIA
  supereight-brane}},  \href{http://arxiv.org/abs/hep-th/9511079}{{\tt
  hep-th/9511079}}.

\bibitem{Seiberg:1996bd}
N.~Seiberg, {\it {Five-dimensional SUSY field theories, nontrivial fixed points
  and string dynamics}},  {\em Phys. Lett. B} {\bf 388} (1996) 753--760,
  [\href{http://arxiv.org/abs/hep-th/9608111}{{\tt hep-th/9608111}}].

\bibitem{Lukas:1998yy}
A.~Lukas, B.~A. Ovrut, K.~S. Stelle, and D.~Waldram, {\it {The Universe as a
  domain wall}},  {\em Phys. Rev. D} {\bf 59} (1999) 086001,
  [\href{http://arxiv.org/abs/hep-th/9803235}{{\tt hep-th/9803235}}].

\bibitem{Lukas:1998tt}
A.~Lukas, B.~A. Ovrut, K.~S. Stelle, and D.~Waldram, {\it {Heterotic M theory
  in five-dimensions}},  {\em Nucl. Phys. B} {\bf 552} (1999) 246--290,
  [\href{http://arxiv.org/abs/hep-th/9806051}{{\tt hep-th/9806051}}].

\bibitem{Lukas:1998uy}
A.~Lukas, B.~A. Ovrut, and D.~Waldram, {\it {Heterotic M theory vacua with
  five-branes}},  {\em Fortsch. Phys.} {\bf 48} (2000) 167--170,
  [\href{http://arxiv.org/abs/hep-th/9903144}{{\tt hep-th/9903144}}].

\bibitem{Ovrut:1999xu}
B.~A. Ovrut, {\it {N=1 supersymmetric vacua in heterotic M theory}},  1, 1999.
\newblock \href{http://arxiv.org/abs/hep-th/9905115}{{\tt hep-th/9905115}}.

\bibitem{Ovrut:2002hi}
B.~A. Ovrut, {\it {Lectures on heterotic M theory}},  in {\em {Theoretical
  Advanced Study Institute in Elementary Particle Physics (TASI 2001): Strings,
  Branes and EXTRA Dimensions}}, 1, 2002.
\newblock \href{http://arxiv.org/abs/hep-th/0201032}{{\tt hep-th/0201032}}.

\bibitem{Mohaupt:2004pr}
T.~Mohaupt and F.~Saueressig, {\it {Dynamical conifold transitions and moduli
  trapping in M-theory cosmology}},  {\em JCAP} {\bf 01} (2005) 006,
  [\href{http://arxiv.org/abs/hep-th/0410273}{{\tt hep-th/0410273}}].

\bibitem{Greene:1995hu}
B.~R. Greene, D.~R. Morrison, and A.~Strominger, {\it {Black hole condensation
  and the unification of string vacua}},  {\em Nucl. Phys. B} {\bf 451} (1995)
  109--120, [\href{http://arxiv.org/abs/hep-th/9504145}{{\tt hep-th/9504145}}].

\bibitem{Chiang:1995hi}
T.-m. Chiang, B.~R. Greene, M.~Gross, and Y.~Kanter, {\it {Black hole
  condensation and the web of Calabi-Yau manifolds}},  {\em Nucl. Phys. B Proc.
  Suppl.} {\bf 46} (1996) 82--95,
  [\href{http://arxiv.org/abs/hep-th/9511204}{{\tt hep-th/9511204}}].

\bibitem{Ibanez:2012zz}
L.~E. Ib{\'{a}}{\~{n}}ez and A.~M. Uranga, {\em {String theory and particle
  physics: An introduction to string phenomenology}}.
\newblock Cambridge University Press, 2012.

\bibitem{Grimm:2018ohb}
T.~W. Grimm, E.~Palti, and I.~Valenzuela, {\it {Infinite Distances in Field
  Space and Massless Towers of States}},  {\em JHEP} {\bf 08} (2018) 143,
  [\href{http://arxiv.org/abs/1802.08264}{{\tt arXiv:1802.08264}}].

\bibitem{Corvilain:2018lgw}
P.~Corvilain, T.~W. Grimm, and I.~Valenzuela, {\it {The Swampland Distance
  Conjecture for Kähler moduli}},  {\em JHEP} {\bf 08} (2019) 075,
  [\href{http://arxiv.org/abs/1812.07548}{{\tt arXiv:1812.07548}}].

\bibitem{Grimm:2019ixq}
T.~W. Grimm, C.~Li, and I.~Valenzuela, {\it {Asymptotic Flux Compactifications
  and the Swampland}},  \href{http://arxiv.org/abs/1910.09549}{{\tt
  arXiv:1910.09549}}.

\bibitem{PopeKK}
C.~Pope, ``{Lectures on Kaluza-Klein}.''
  \url{http://people.tamu.edu/~c-pope/ihplec.pdf}.

\bibitem{Blumenhagen:2019vgj}
R.~Blumenhagen, M.~Brinkmann, and A.~Makridou, {\it {Quantum Log-Corrections to
  Swampland Conjectures}},  {\em JHEP} {\bf 02} (2020) 064,
  [\href{http://arxiv.org/abs/1910.10185}{{\tt arXiv:1910.10185}}].

\bibitem{Bandos:2018gjp}
I.~Bandos, F.~Farakos, S.~Lanza, L.~Martucci, and D.~Sorokin, {\it
  {Three-forms, dualities and membranes in four-dimensional supergravity}},
  {\em JHEP} {\bf 07} (2018) 028, [\href{http://arxiv.org/abs/1803.01405}{{\tt
  arXiv:1803.01405}}].

\bibitem{Sugimoto:1999tx}
S.~Sugimoto, {\it {Anomaly cancellations in type I D-9 - anti-D-9 system and
  the USp(32) string theory}},  {\em Prog. Theor. Phys.} {\bf 102} (1999)
  685--699, [\href{http://arxiv.org/abs/hep-th/9905159}{{\tt hep-th/9905159}}].

\bibitem{Klebanov:2000hb}
I.~R. Klebanov and M.~J. Strassler, {\it {Supergravity and a confining gauge
  theory: Duality cascades and chi SB resolution of naked singularities}},
  {\em JHEP} {\bf 08} (2000) 052,
  [\href{http://arxiv.org/abs/hep-th/0007191}{{\tt hep-th/0007191}}].

\bibitem{Calderon-Infante:2020dhm}
J.~Calder\'on-Infante, A.~M. Uranga, and I.~Valenzuela, {\it {The Convex Hull
  Swampland Distance Conjecture and Bounds on Non-geodesics}},  {\em JHEP} {\bf
  03} (2021) 299, [\href{http://arxiv.org/abs/2012.00034}{{\tt
  arXiv:2012.00034}}].

\bibitem{Cribiori:2021gbf}
N.~Cribiori, D.~Lust, and M.~Scalisi, {\it {The gravitino and the swampland}},
  {\em JHEP} {\bf 06} (2021) 071, [\href{http://arxiv.org/abs/2104.08288}{{\tt
  arXiv:2104.08288}}].

\bibitem{Castellano:2021yye}
A.~Castellano, A.~Font, A.~Herraez, and L.~E. Ib\'a\~nez, {\it {A Gravitino
  Distance Conjecture}},  \href{http://arxiv.org/abs/2104.10181}{{\tt
  arXiv:2104.10181}}.

\bibitem{Bonnefoy:2019nzv}
Q.~Bonnefoy, L.~Ciambelli, D.~L\"ust, and S.~L\"ust, {\it {Infinite Black Hole
  Entropies at Infinite Distances and Tower of States}},  {\em Nucl. Phys. B}
  {\bf 958} (2020) 115112, [\href{http://arxiv.org/abs/1912.07453}{{\tt
  arXiv:1912.07453}}].

\bibitem{Klebanov:2000nc}
I.~R. Klebanov and A.~A. Tseytlin, {\it {Gravity duals of supersymmetric SU(N)
  x SU(N+M) gauge theories}},  {\em Nucl. Phys.} {\bf B578} (2000) 123--138,
  [\href{http://arxiv.org/abs/hep-th/0002159}{{\tt hep-th/0002159}}].

\bibitem{Franco:2005fd}
S.~Franco, A.~Hanany, and A.~M. Uranga, {\it {Multi-flux warped throats and
  cascading gauge theories}},  {\em JHEP} {\bf 09} (2005) 028,
  [\href{http://arxiv.org/abs/hep-th/0502113}{{\tt hep-th/0502113}}].

\bibitem{GarciaEtxebarria:2006aq}
I.~Garcia-Etxebarria, F.~Saad, and A.~M. Uranga, {\it {Quiver gauge theories at
  resolved and deformed singularities using dimers}},  {\em JHEP} {\bf 06}
  (2006) 055, [\href{http://arxiv.org/abs/hep-th/0603108}{{\tt
  hep-th/0603108}}].

\bibitem{Franco:2005zu}
S.~Franco, A.~Hanany, F.~Saad, and A.~M. Uranga, {\it {Fractional branes and
  dynamical supersymmetry breaking}},  {\em JHEP} {\bf 01} (2006) 011,
  [\href{http://arxiv.org/abs/hep-th/0505040}{{\tt hep-th/0505040}}].

\bibitem{Hanany:2005ve}
A.~Hanany and K.~D. Kennaway, {\it {Dimer models and toric diagrams}},
  \href{http://arxiv.org/abs/hep-th/0503149}{{\tt hep-th/0503149}}.

\bibitem{Franco:2005rj}
S.~Franco, A.~Hanany, K.~D. Kennaway, D.~Vegh, and B.~Wecht, {\it {Brane dimers
  and quiver gauge theories}},  {\em JHEP} {\bf 01} (2006) 096,
  [\href{http://arxiv.org/abs/hep-th/0504110}{{\tt hep-th/0504110}}].

\bibitem{Kennaway:2007tq}
K.~D. Kennaway, {\it {Brane Tilings}},  {\em Int. J. Mod. Phys.} {\bf A22}
  (2007) 2977--3038, [\href{http://arxiv.org/abs/0706.1660}{{\tt
  arXiv:0706.1660}}].

\bibitem{Franco:2004jz}
S.~Franco, Y.-H. He, C.~Herzog, and J.~Walcher, {\it {Chaotic duality in string
  theory}},  {\em Phys. Rev.} {\bf D70} (2004) 046006,
  [\href{http://arxiv.org/abs/hep-th/0402120}{{\tt hep-th/0402120}}].

\bibitem{Retolaza:2015sta}
A.~Retolaza, A.~M. Uranga, and A.~Westphal, {\it {Bifid Throats for Axion
  Monodromy Inflation}},  {\em JHEP} {\bf 07} (2015) 099,
  [\href{http://arxiv.org/abs/1504.02103}{{\tt arXiv:1504.02103}}].

\bibitem{McAllister:2008hb}
L.~McAllister, E.~Silverstein, and A.~Westphal, {\it {Gravity Waves and Linear
  Inflation from Axion Monodromy}},  {\em Phys. Rev.} {\bf D82} (2010) 046003,
  [\href{http://arxiv.org/abs/0808.0706}{{\tt arXiv:0808.0706}}].

\bibitem{Silverstein:2008sg}
E.~Silverstein and A.~Westphal, {\it {Monodromy in the CMB: Gravity Waves and
  String Inflation}},  {\em Phys. Rev.} {\bf D78} (2008) 106003,
  [\href{http://arxiv.org/abs/0803.3085}{{\tt arXiv:0803.3085}}].

\bibitem{Kaloper:2008fb}
N.~Kaloper and L.~Sorbo, {\it {A Natural Framework for Chaotic Inflation}},
  {\em Phys. Rev. Lett.} {\bf 102} (2009) 121301,
  [\href{http://arxiv.org/abs/0811.1989}{{\tt arXiv:0811.1989}}].

\bibitem{Marchesano:2014mla}
F.~Marchesano, G.~Shiu, and A.~M. Uranga, {\it {F-term Axion Monodromy
  Inflation}},  {\em JHEP} {\bf 09} (2014) 184,
  [\href{http://arxiv.org/abs/1404.3040}{{\tt arXiv:1404.3040}}].

\bibitem{Hebecker:2014eua}
A.~Hebecker, S.~C. Kraus, and L.~T. Witkowski, {\it {D7-Brane Chaotic
  Inflation}},  {\em Phys. Lett.} {\bf B737} (2014) 16--22,
  [\href{http://arxiv.org/abs/1404.3711}{{\tt arXiv:1404.3711}}].

\bibitem{McAllister:2014mpa}
L.~McAllister, E.~Silverstein, A.~Westphal, and T.~Wrase, {\it {The Powers of
  Monodromy}},  {\em JHEP} {\bf 09} (2014) 123,
  [\href{http://arxiv.org/abs/1405.3652}{{\tt arXiv:1405.3652}}].

\end{thebibliography}\endgroup

\end{document}